\documentclass[twocolumn,aps,prd,epsfig,nofootinbib]{revtex4}
\usepackage{epsfig}
\usepackage{color}
\usepackage{bm}
\usepackage{subfigure}
\usepackage{leftidx}
\usepackage{slashed}
\usepackage{epstopdf}

\newcommand{\nn}{\nonumber}

\begin{document}

\vspace*{-15mm}
\hfill
\begin{flushright}
MADPH-11-1570
\end{flushright}

\title{{Discovery in Drell-Yan Processes at the LHC}}

\author{Cheng-Wei Chiang$^{1,2,3,4}$, Neil D. Christensen$^{4,5}$, Gui-Jun Ding$^6$, and Tao Han$^{4,5}$}

\affiliation{%
\bigskip
$^1$ Department of Physics and Center for Mathematics and Theoretical Physics,
National Central University, Chungli, Taiwan 32001, ROC \\
$^2$ Institute of Physics, Academia Sinica, Taipei, Taiwan 11925, ROC \\
$^3$ Physics Division, National Center for Theoretical Sciences, Hsinchu, Taiwan 30043, ROC\\
$^4$ Department of Physics, University of Wisconsin-Madison, Madison, WI 53706, USA \\
$^5$ Department of Physics $\&$ Astronomy, University of Pittsburgh, Pittsburgh, PA 15260, USA \\
$^6$ Department of Modern Physics, University of Science and Technology of China, Hefei, Anhui 230026, PRC}

\begin{abstract}
We study the Drell-Yan process mediated by a new bosonic resonance at the LHC.  The bosons of spin-0, 1, and 2 with the most general leading-order couplings to Standard Model fermions and gluons are considered, which provide a model-independent formulation for future exploration of the resonance properties, such as its spin, mass and couplings.
In the case of neutral resonances,
we demonstrate how the shapes of the kinematical distributions
change as one varies the chiral couplings of the quarks and leptons,
and show how to analyze the couplings by making use of the
forward-backward asymmetry.  In the case of charged resonances, we
propose a novel technique to effectively reconstruct the angular
distribution in the center-of-mass frame, to a large extent avoiding the two-fold
ambiguity due to the missing neutrino.
Similar to the case of a neutral resonance, the spin information of the resonance can be extracted unambiguously, and
chiral couplings and the asymmetries can be explored in a statistical manner.
With the current LHC data, we present bounds on the mass and cross section times branching fraction of the new resonance and estimate the future reach.

\end{abstract}

\pacs{}
\maketitle

%
%

\section{Introduction
\label{sec:intro}}

With the start of the LHC experiments, we have entered a new era of high-energy physics that directly probes Nature at the TeV scale.  Depending on the underlying theory, new particles of different kinds may lead to novel and distinctive signatures at the collider. 
In addition to the highly anticipated discovery of the origin of electroweak symmetry breaking, 
we will likely also discover other new resonances associated with this scale, presumably through the classic Drell-Yan (DY) process with a striking signal at around the TeV scale.  If such a new resonance is indeed observed, it is important to determine many of its properties, such as the spin $J$, fermionic chiral couplings, and so on, in addition to its mass, width and electric charge.
In this work, we consider new spin-0, 1, and 2 resonances that can contribute to the $s$-channel Drell-Yan processes\footnote{There have been many studies of resonances that can contribute to the Drell-Yan process.  See, for example, 
\cite{Nakamura:2010zzi-ZpWp,Davoudiasl:2000wi,Allanach:2000nr,del Aguila:1986ez,Barger:1986hd,Baur:1986zt,Langacker:1991pg,delAguila:1993ym,Abreu:1994ria,Riemann:1996fk,Leike:1998wr,Rizzo:2003ug,Dittmar:2003ir,Carena:2004xs,Burikham:2004su,Rizzo:2006nw,Djouadi:2007eg,Agashe:2007ki,Fuks:2007gk,Langacker:2008yv,Petriello:2008zr,Coriano:2008wf,Davoudiasl:2008hx,Petriello:2008pu,Gershtein:2008bf,Rizzo:2009pu,Osland:2009tn,Erler:2009jh,Li:2009xh,Salvioni:2009mt,Diener:2009vq,Langacker:2009su,delAguila:2010mx,Accomando:2010fz,Abe:2011qe,Keung:1983uu,Frere:1990qm,Sullivan:2002jt,Birkedal:2004au,Frank:2010cj,Gopalakrishna:2010xm,Schmaltz:2010xr,Grojean:2011vu,Accomando:2011xi,Langacker:1984dc,Cvetic:1995zs,Cheng:2002ab,Han:2003wu,Han:2005ru,Agashe:2006hk,SekharChivukula:2007gi,Agashe:2008jb,Alwall:2008ag,Cata:2009iy,Barbieri:2009tx,Wang:2011uq}
, where 
\cite{Davoudiasl:2000wi,Allanach:2000nr,del Aguila:1986ez,Barger:1986hd,Baur:1986zt,Langacker:1991pg,delAguila:1993ym,Abreu:1994ria,Riemann:1996fk,Leike:1998wr,Rizzo:2003ug,Dittmar:2003ir,Carena:2004xs,Burikham:2004su,Rizzo:2006nw,Djouadi:2007eg,Agashe:2007ki,Fuks:2007gk,Langacker:2008yv,Petriello:2008zr,Coriano:2008wf,Davoudiasl:2008hx,Petriello:2008pu,Gershtein:2008bf,Rizzo:2009pu,Osland:2009tn,Erler:2009jh,Li:2009xh,Salvioni:2009mt,Diener:2009vq,Langacker:2009su,delAguila:2010mx,Accomando:2010fz,Abe:2011qe}
 deal with a neutral boson,  
 \cite{Keung:1983uu,Frere:1990qm,Sullivan:2002jt,Birkedal:2004au,Frank:2010cj,Gopalakrishna:2010xm,Schmaltz:2010xr,Grojean:2011vu,Accomando:2011xi,Agashe:2008jb}
  deal with a charged boson and 
  \cite{Langacker:1984dc,Cvetic:1995zs,Cheng:2002ab,Han:2003wu,Han:2005ru,Agashe:2006hk,SekharChivukula:2007gi,Alwall:2008ag,Cata:2009iy,Barbieri:2009tx,Wang:2011uq}%
   deal with both.}.
In view of many possible models predicting such resonances, we keep the couplings between the standard model (SM) fermions and the resonances as general as allowed by symmetries. 

For a new neutral resonance, the final state involves two charged leptons whose momentum information can be fully registered by the detector.  Therefore, with a sufficient rate, the particle mass can be readily determined from the invariant mass distribution.  By boosting to the center-of-mass (CM) frame of the two leptons and noting that the
boost direction preferably coincides with the momentum of the colliding quark, one can study the angular distribution of the leptons to extract the spin information of the resonance.

The case of a charged resonance is more complicated.  This is because the final state contains a charged lepton and the associated neutrino, thus missing energy is involved in such events.  The resonance mass can be best determined from the Jacobian peak in the transverse mass distribution.  However, there is a difficulty in finding the correct CM frame of the charged lepton and the neutrino.  Even if one assumes that the resonance mass has been measured, there are generally two possible solutions for the longitudinal momentum of the neutrino.  We develop a novel technique to effectively construct the angular distribution for the charged lepton in the CM frame in a statistical manner, and show how the spin and chiral couplings of the resonance can be extracted in a similar fashion as in the case of neutral resonances.  

This paper is organized as follows.  In Section~\ref{sec:DYgeneral},
we classify types of new resonances that can mediate $s$-channel
Drell-Yan processes at the LHC, and discuss some of the general
features of these processes.  In addition, we discuss the most
general interactions between SM fermions and the new resonances
allowed by the Lorentz and gauge invariances, with the corresponding
Feynman rules given in Appendix~\ref{sec:feynman rules}.
Section~\ref{sec:v} is devoted to the discussions of the Drell-Yan
process mediated by a neutral resonance.
The
Drell-Yan process associated with a charged resonance is analyzed in
Section~\ref{sec:vprime}.  The current results of the LHC are used
to place a bound on the masses and
couplings of these new resonances in Section~\ref{sec:Search}. The
findings of this work are summarized in Section~\ref{sec:summary}.
Appendix~\ref{app:FeynRules} contains details of the FeynRules
implementation of these new interactions.  
Appendix C gives a short review of the Wigner $d^j_{m,m'}$ functions that are useful for the helicity amplitudes of our calculations.

%
%

\section{General Interactions of new Drell-Yan resonances
\label{sec:DYgeneral}}

We will concentrate exclusively on color-singlet 
neutral-current and charged-current $s$-channel resonances contributing to the Drell-Yan processes at the LHC
\begin{equation}
pp \to \ell^+ \ell^- X\quad {\rm and} \quad  \ell^\pm \nu X ~,
\end{equation}
where $\ell$ generically denotes either an electron or a muon, $\nu$ denotes a neutrino or an antineutrino and
$X$ the inclusive hadronic remnants. We consider the most general
couplings for leading-order operators allowed by certain symmetries
for such a new particle.  This particle must be a color singlet and 
have integer spin.  We concentrate
on the possibilities of a scalar ($S$), vector ($V$) and traceless
symmetric second-rank tensor ($T$), although other spins are
possible in principle.  There are two cases for the electrical
charge of the boson.  In the case of neutral-current processes, this
boson (generically denoted by $R_{0}$) must be neutral whereas for
the charged-current processes, this boson (generically denoted by
$R_{c}$) must have charge $\pm1$. In Table~\ref{tab:quantumnumber},
we summarize the properties of these new bosons along with the
processes to which they contribute.

%
\begin{table}[hptb]
\begin{tabular}{ccccc}\hline\hline
Notation~~ &  $|Q_e|$ & $~~~J$ & $~~~$ Partonic processes
\\
\hline
$R_{0}$  &  0 & $~~~~$ 0,\,1,\,2 & $~~~~u\bar{u},d\bar{d},gg\rightarrow\ell^+\ell^-$ \\
$R_{c}$ &  1 & $~~~~$ 0,\,1,\,2 & $~~~~~~~u\overline{d}\rightarrow\ell^+\nu$,\,$d\overline{u}\rightarrow \ell^-\overline{\nu}$ \\
\hline\hline
\end{tabular}
\caption{Resonance particles, their quantum numbers, and $s$-channel Drell-Yan processes. 
$Q_e$ and $J$ represent their electric charge and spin respectively.  
\label{tab:quantumnumber}}
\end{table}

We now write down the most general Lagrangian between these new
bosons and the SM fermions and gluons allowed by Lorentz, quantum
chromodynamic and electromagnetic gauge invariance. In each case, we
only include the leading effective terms which are either
dimension-4 or dimension-5 operators.  Furthermore, we drop all
terms which vanish when the masses of the initial-state and
final-state particles are taken to zero.

\subsection{Spin-0 states}

 We begin with the neutral 
  scalar boson $S$ which can have the following Lagrangian
\begin{equation}
{\cal L}_{S}
= \overline{f}_{i}\left(g^f_{Sij}+ig^f_{Pij}\gamma_5\right)f_{j} S
-\frac{1}{4} \frac{g_S^g}{\Lambda} F_{\mu\nu}^a F^{a\mu\nu}S ,
\end{equation}
where $f$ can be either a quark or a lepton.
The indices $i$ and $j$ run over generations, and the generation matrices $g^f_{Sij}=(g^f_S)_{ij}$ and $g^f_{Pij}=(g^f_P)_{ij}$ are required to be Hermitian by the Hermiticity of the Lagrangian.
 $F_{\mu\nu}^a$ denotes the gluonic field strength tensor. Here and henceforth, $\Lambda$ denotes the cutoff scale of the effective interactions, which should be at least at the order of the resonance mass or higher.
%
%
For the charged scalar boson $S^{\pm}$, we have
\begin{eqnarray}
{\cal L}_{S^\pm}
&=&
\overline{u}_i\left(h^q_{Sij}
+ ih^q_{Pij}\gamma_5\right)d_{j}S^+
+ \mbox{h.c.}
\nonumber\\
&&
+ \overline{\nu}_i\left(h^{\ell}_{Sij}
+ ih^{\ell}_{Pij}\gamma_5\right)\ell_{j}S^+ + \mbox{h.c.} ,
\end{eqnarray}
where $h^f_{Sij}=(h_S^f)_{ij}$ and $h^f_{Pij}=(h_P^f)_{ij}$ are allowed to be general complex matrices.

\subsection{Spin-1 states}

Using the same notation, the most general Lagrangian for the neutral vector boson $V_\mu$ is
\begin{equation}
{\cal L}_{V}
= \overline{f}_{i}\gamma^{\mu}\left(g^f_{Vij}+g^f_{Aij}\gamma_5\right)f_{j}V_{\mu} ,
\end{equation}
where $g^f_{Vij}=(g^f_V)_{ij}$ and $g^f_{Aij}=(g^f_A)_{ij}$ are
required to be Hermitian matrices.\footnote{Our convention is fixed
with respect to the SM couplings as $g_{ZV}^{f} =
g_{Z}({1\over 2}T_{3}^{f} -
Q^{f}s_{w}^{2}),\ g_{ZA}^{f}
= -{1\over 2}T_{3}^{f} g_{Z}, $
where $g_Z=g/c_w$, $g$ is the weak coupling and  $s_w$ and  $c_w$ are the sine and cosine of the Weinberg angle.
The left- and right-chiral couplings are related to the vector and axial-vector couplings as
$g_{R_XL}=g_{R_XV}-g_{R_XA},\ g_{R_XR}=g_{R_XV}+g_{R_XA}$. }
We have dropped interactions of the neutral vector boson $V$ with gluons since they do not contribute to this Drell-Yan
process.
   The charged vector boson has the Lagrangian
\begin{eqnarray}
{\cal L}_{V^\pm}
&=&
\overline{u}_i\gamma^{\mu}\left(h^q_{Vij}+h^q_{Aij}\gamma_5\right)d_{j}V^+_{\mu}
+\mbox{h.c.}\nonumber\\
&&
+ \overline{\nu}_i\gamma^{\mu}\left(h^{\ell}_{Vij}+h^{\ell}_{Aij}\gamma_5\right)
\ell_{j}V^+_{\mu}
+\mbox{h.c.} ,
\end{eqnarray}
where $h^f_{Vij}=(h_V^f)_{ij}$ and $h^f_{Aij}=(h_A^f)_{ij}$ are allowed to be general
complex matrices.

\subsection{Spin-2 states}

The neutral tensor Lagrangian is given by
\begin{eqnarray}
{\cal L}_T&=&
\frac{i}{\Lambda} \bigg[
\overline{f}_i\left(g^{f}_{Tij}-g^{f}_{ATij}\gamma_5\right)
\left(\gamma^{\mu}\partial^{\nu}f_{j}+\gamma^{\nu}\partial^{\mu}f_{j}\right) \nonumber\\
&&\nonumber\\
&&
- \left(\partial^{\mu}\overline{f}_i\gamma^{\nu} + \partial^{\nu}\overline{f}_i\gamma^{\mu}\right)
\left(g^{f \dagger}_{Tij}+g^{f \dagger}_{ATij}\gamma_5\right)f_{j}
\bigg]  T_{\mu\nu}
\nonumber\\
&&
-\frac{1}{4} \frac{g_T^g}{\Lambda} F_{\mu\alpha}^a F_\beta^{a\mu} T^{\alpha\beta} ,
\end{eqnarray}
where the couplings $g^f_{Tij}=(g^{f}_T)_{ij}$ and $g^f_{ATij}=(g^{f}_{AT})_{ij}$ are general $3\times3$ complex matrices.  Note that a dimension-4 operator between SM fermions and neutral tensor particle is not allowed since it would be proportional to the trace of the tensor which we have assumed to be traceless.  
Finally, the charged tensor has the interaction Lagrangian
as follows
\begin{eqnarray}
{\cal L}_{T^\pm}\hspace{-0.05in}&=&
\frac{i}{\Lambda} \bigg[
\overline{u}_i\left(h^{q}_{Tij}-h^{q}_{ATij}\gamma_5\right)
\left(\gamma^{\mu}\partial^{\nu}d_{j}+\gamma^{\nu}\partial^{\mu}d_{j}\right)
\nn \\
&& - (\partial^{\mu}\overline{u}_i\gamma^{\nu}
+\partial^{\nu}\overline{u}_i\gamma^{\mu})
\left(\tilde{h}^{q}_{Tij}+\tilde{h}^{q}_{ATij}\gamma_5\right)d_{j}
 \bigg] T^+_{\mu\nu}
 \nn \\
 && + \frac{i}{\Lambda} \bigg[
\overline{\nu}_i\left(h^{\ell}_{Tij}-h^{\ell}_{ATij}\gamma_5\right)
\left(\gamma^{\mu}\partial^{\nu}\ell_{j}+\gamma^{\nu}\partial^{\mu}\ell_{j}\right)
\nn \\
&& - (\partial^{\mu}\overline{\nu}_i\gamma^{\nu}
+\partial^{\nu}\overline{\nu}_i\gamma^{\mu})
\left(\tilde{h}^{\ell}_{Tij}+\tilde{h}^{\ell}_{ATij}\gamma_5\right)\ell_{j}
 \bigg] T^+_{\mu\nu}
 \nn \\
 &&
 + \mbox{h.c.} ,
\end{eqnarray}
where $h^f_{Tij}=(h_T^f)_{ij}$, $h^f_{AFij}=(h_{AT}^f)_{ij}$, $\tilde{h}^f_{Tij}=(\tilde{h}_T^f)_{ij}$ and $\tilde{h}^f_{ATij}=(\tilde{h}_{AT}^f)_{ij}$ are general complex matrices.  The corresponding Feynman rules have been worked out and can be found in Appendix~\ref{sec:feynman rules}.

%
%

\begin{figure*}[!tb]
\begin{tabular}{p{0.8in}p{0.8in}p{0.8in}p{0.8in}p{0.8in}p{0.8in}p{0.8in}p{0.8in}}
\multicolumn{2}{c}{\includegraphics[scale=.43]{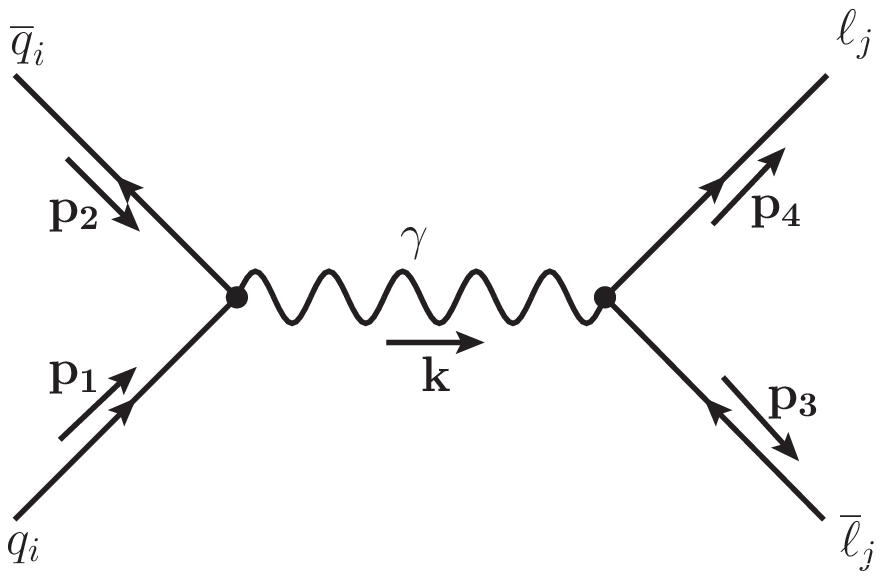}} &
\multicolumn{2}{c}{\includegraphics[scale=.43]{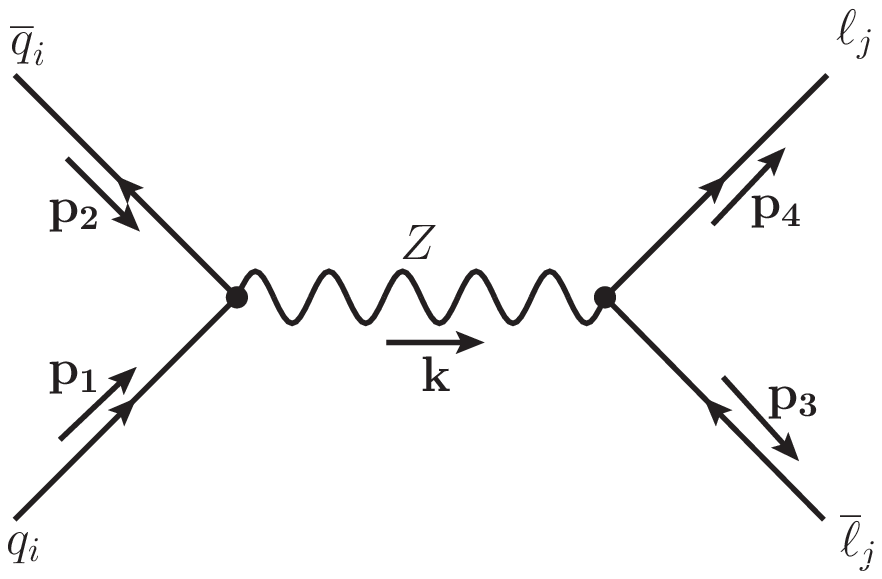}} &
\multicolumn{2}{c}{\includegraphics[scale=.43]{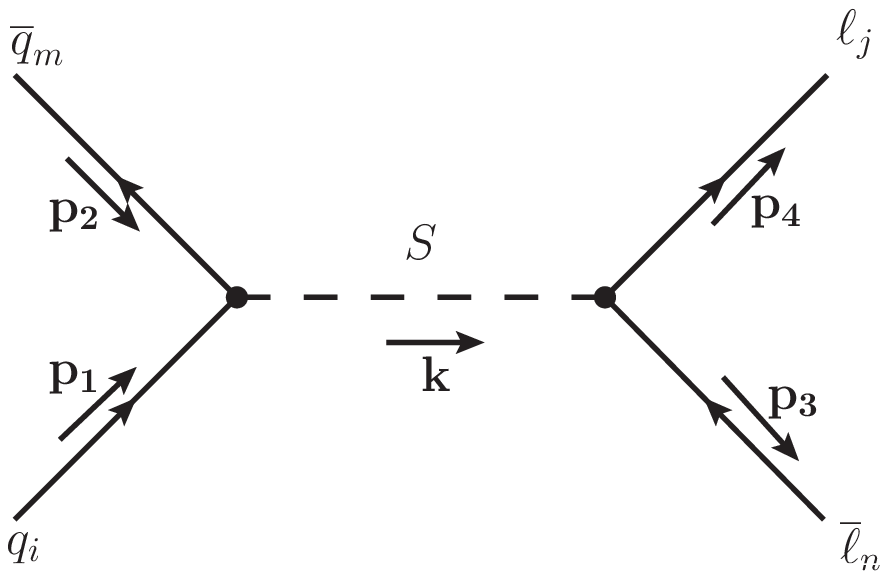}} &
\multicolumn{2}{c}{\includegraphics[scale=.43]{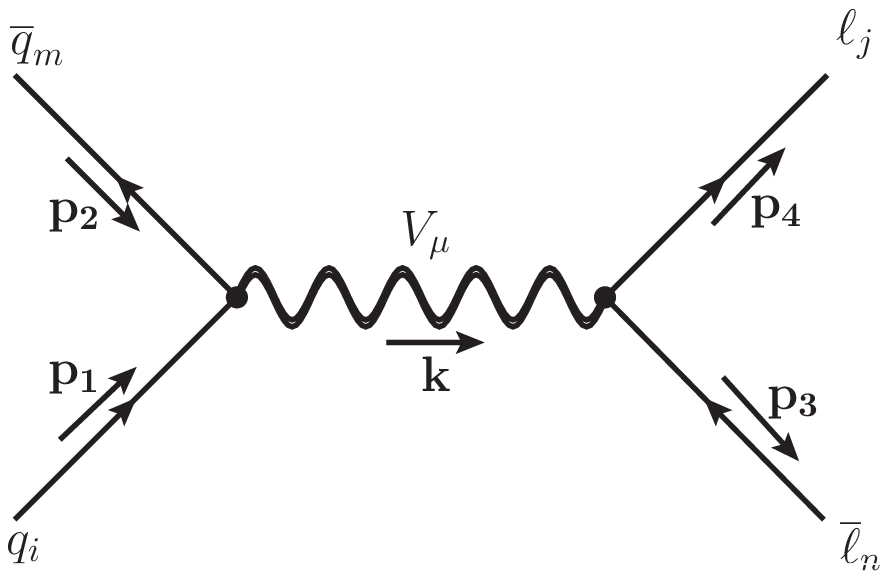}} \\
\multicolumn{2}{c}{($\gamma$)} &
\multicolumn{2}{c}{($Z$)} &
\multicolumn{2}{c}{($S$)} &
\multicolumn{2}{c}{($V$)} \\

&
\multicolumn{2}{c}{\includegraphics[scale=.43]{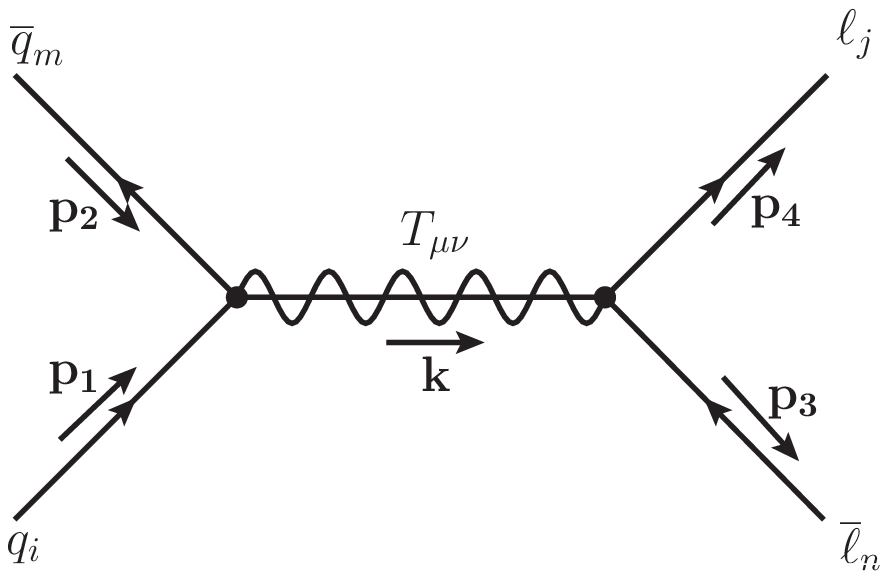}} &
\multicolumn{2}{c}{\includegraphics[scale=.43]{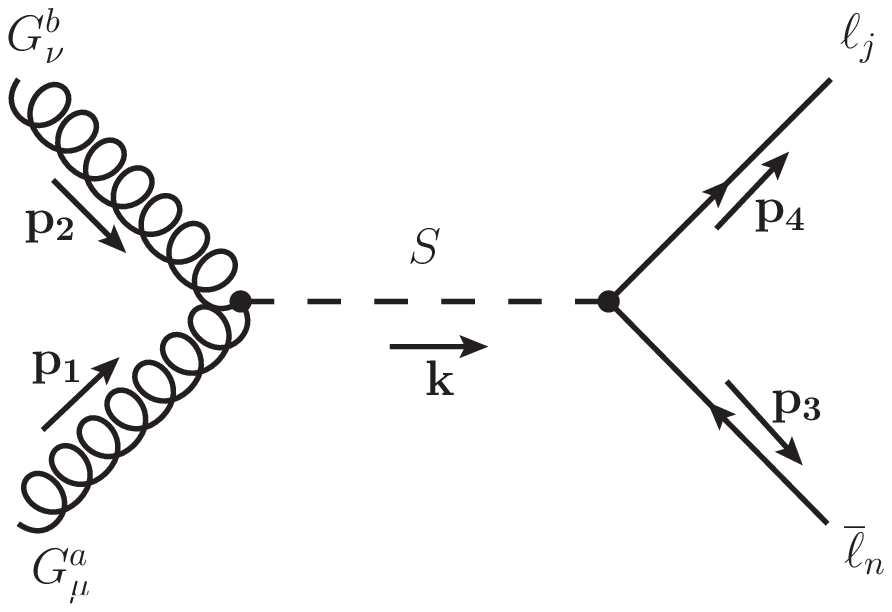}} &
\multicolumn{2}{c}{\includegraphics[scale=.43]{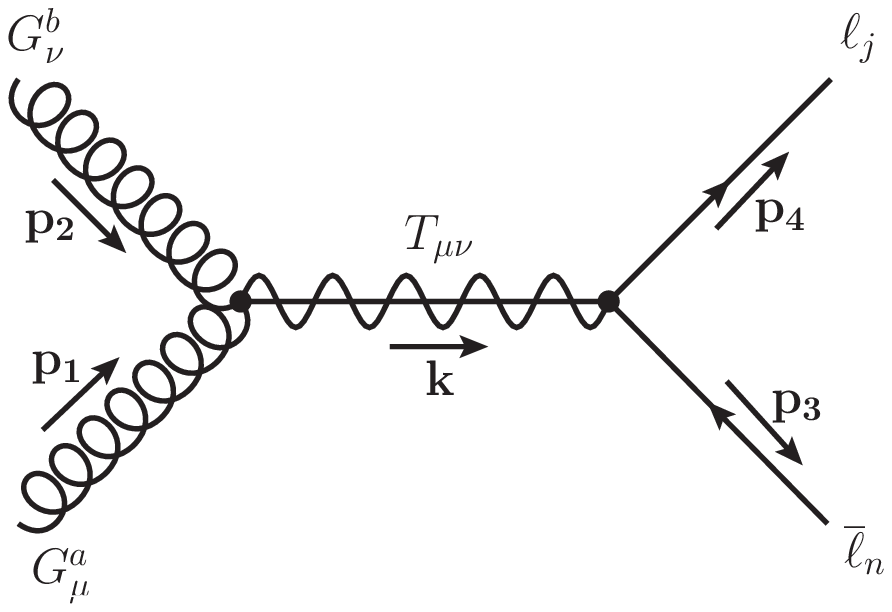}} \\
&
\multicolumn{2}{c}{($T$)} &
\multicolumn{2}{c}{($gS$)} &
\multicolumn{2}{c}{($gT$)}
\end{tabular}
\caption{The $s$-channel Feynman diagrams responsible for $pp \rightarrow \ell^+ \ell^- X$.  Contributions to $\bar{q}q\rightarrow\ell^+\ell^-$ are from a photon ($\gamma$), a $Z$ boson ($Z$), a new scalar particle ($S$), a new vector particle ($V$) and a new tensor particle ($T$).  Contributions to $gg\rightarrow\ell^+\ell^-$ are from a new scalar field ($gS$) and a new tensor field ($gT$).\label{scatter_diagram_Z} }
\end{figure*}

\section{Neutral Boson Resonances \label{sec:v}}

Diagrams for the Drell-Yan process $pp\rightarrow\ell^+\ell^-X$
mediated by a neutral boson at tree level are shown in
Fig.~\ref{scatter_diagram_Z}, including the SM diagrams with $\gamma/Z$ exchanges.  The
corresponding helicity amplitudes are listed in Table~\ref{tab:helam_neutral} where
\begin{equation}
D_X = s-M_X^2+iM_X\Gamma_X .
\end{equation}

\begin{table*}[tb]
\begin{center}
\renewcommand{\arraystretch}{2.5}
\renewcommand{\tabcolsep}{0.25in}
\begin{tabular}{|ll|}

\multicolumn{2}{c}{\boldmath $q_{i}(\lambda) \bar q_{m}(-\lambda)\rightarrow \ell_{j}(\lambda') \bar \ell_{n}(-\lambda')$} \\
\hline\hline

$\mathcal{M}^{\lambda\lambda'}_\gamma=\mathcal{N}^{\lambda\lambda'}_\gamma d^{1}_{1,\lambda\lambda'}$  & $\mathcal{N}^{\lambda\lambda'}_\gamma=8\pi\alpha\lambda\lambda' Q_q$\\

$\mathcal{M}^{\lambda\lambda'}_Z=\mathcal{N}^{\lambda\lambda'}_Z d^{1}_{1,\lambda\lambda'}$  & $\mathcal{N}^{\lambda\lambda'}_Z=-\frac{2s}{D_Z}(\lambda g_{ZV}^q+g_{ZA}^q)(\lambda' g_{ZV}^{\ell}+g_{ZA}^{\ell})$\\

$\mathcal{M}^{\lambda\lambda'}_V=\mathcal{N}^{\lambda\lambda'}_V d^{1}_{1,\lambda\lambda'}$ & $\mathcal{N}^{\lambda\lambda'}_V=-\frac{2s}{D_V}(\lambda g^{q}_{V}+g^{q}_{A})_{mi}(\lambda'g^{\ell}_{V}+g^{\ell}_{A})_{jn}$ \\

$\mathcal{M}^{\lambda\lambda'}_T=\mathcal{N}^{\lambda\lambda'}_{T2}\,d^{2}_{1,\lambda\lambda'}+\mathcal{N}^{\lambda\lambda'}_{T1}\,d^{1}_{1,\lambda\lambda'}$ & $\mathcal{N}^{\lambda\lambda'}_{T2}=-\frac{2\lambda\lambda's^2}{\Lambda^2 D_T}\mathcal{H}_+(g^{q}_{T}+\lambda g^{q}_{AT})_{mi}\mathcal{H}_+(g^{\ell}_T+\lambda'g^{\ell}_{AT})_{jn}$\\

& $\mathcal{N}^{\lambda\lambda'}_{T1}=\frac{2\lambda\lambda's^2 (s-M_T^2)}{\Lambda^2 M^2_T D_T}\mathcal{H}_-(g^{q}_{T}+\lambda g^{q}_{AT})_{mi}\,\mathcal{H}_-(g^{\ell}_T+\lambda'g^{\ell}_{AT})_{jn}$\\
\hline

\multicolumn{2}{c}{\boldmath $g^{a}(\lambda)g^{b}(-\lambda)\rightarrow  \ell_{j}(\lambda') \bar \ell_n(-\lambda')$} \\
\hline\hline
$\mathcal{M}^{\lambda\lambda'}_{gT}=\mathcal{N}^{\lambda\lambda'}_{gT}d^{2}_{1,\lambda\lambda'}$ & $\mathcal{N}^{\lambda\lambda'}_{gT}=-\frac{\lambda\lambda'g^{g}_T s^2}{2\Lambda^2 D_T}\mathcal{H}_+(g^{\ell}_{Tjn}+\lambda'g^{\ell}_{ATjn})\delta^{ab}$\\
\hline
%
%
\multicolumn{2}{c}{\boldmath $q_{i}(\lambda) \bar q_{m}(\lambda)\rightarrow \ell_{j}(\lambda') \bar \ell_{n}(\lambda')$} \\
\hline\hline
$\mathcal{M}^{\lambda\lambda'}_S=\mathcal{N}^{\lambda\lambda'}_S d^{0}_{0,0}$ & $\mathcal{N}^{\lambda\lambda'}_S=\frac{s}{D_S}(i\lambda g^{q}_{Smi}-g^{q}_{Pmi})(i\lambda'g^{\ell}_{S}+g^{\ell}_{P})_{jn}$\\
\hline
\multicolumn{2}{c}{\boldmath $g^{a}(\lambda)g^{b}(\lambda)\rightarrow  \ell_{j}(\lambda') \bar \ell_n(\lambda')$} \\
\hline\hline
$\mathcal{M}^{\lambda\lambda'}_{gS}=\mathcal{N}^{\lambda\lambda'}_{gS}d^{0}_{0,0}$ & $\mathcal{N}^{\lambda\lambda'}_{gS}=\frac{g^{g}_{S}s^{3/2}}{2\Lambda D_S}(\lambda'g^{\ell}_{S}-ig^{\ell}_{P})_{jn}\delta^{ab}$\\

\hline
\end{tabular}
\caption{\label{tab:helam_neutral}Helicity scattering amplitudes for
the parton-level processes.  The amplitudes correspond to the
diagrams in Fig.~\ref{scatter_diagram_Z}.  The particles in the $s$-channel exchange are labelled by subscripts ($\gamma$, $Z$, $S$, $V$
and $T$) while $gS$ and $gT$ indicate that the initial states are
gluons with scalar and tensor exchange, respectively.  $\lambda$ and
$\lambda'$ are the helicities and we define $\mathcal{H}_\pm(M)=M\pm
M^\dagger$.}
\end{center}
\end{table*}

\noindent
We have expressed the scattering amplitudes in terms of the Wigner $d^{j}_{m,m'}$ functions,
where $j$ is the total angular momentum and $m$ and $m'$ are the difference of the helicities of the initial-state and final-state particles, respectively \cite{Nakamura:2010zzi}.  A short review of the $d^j_{m,m'}$ functions can be found in Appendix \ref{app:d functions}.
For $s$-channel scalar resonances, the initial state particles must have the same helicities to conserve angular momentum.  The same is true for the final state particles.  For this reason, we find that only the helicity combinations $(\lambda,\lambda)\rightarrow(\lambda',\lambda')$ are nonzero.  On the other hand, for $s$-channel vector and tensor interactions, angular momentum is not sufficient to determine the helicity combinations of the external particles.  
However, the combined effect of the masslessness of the external fields and the properties of the interaction vertices only allows opposite helicities for the incoming particles and also for the outgoing particles
 $(\lambda,-\lambda)\rightarrow(\lambda',-\lambda')$. 

\subsection{Invariant mass spectrum
\label{sec:v-invmass}}

\begin{figure*}[tb]
\begin{center}
\begin{tabular}{cccccc}
\raisebox{0.35in}[0pt]{
\begin{tabular}{c}
$\phi_\ell=\pi$ \\
$g_\ell\propto-1$ 
\end{tabular}} &
\includegraphics[scale=0.85]{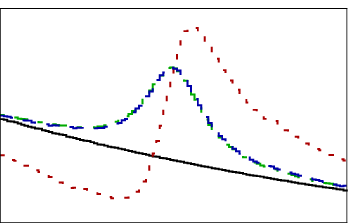} &
\includegraphics[scale=0.85]{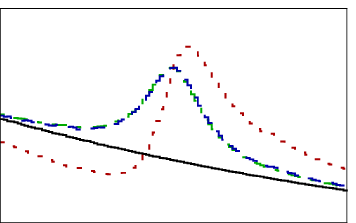} &
\includegraphics[scale=0.85]{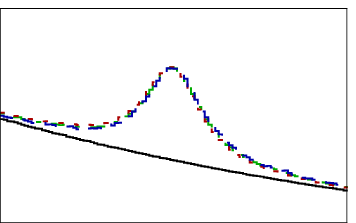} &
\includegraphics[scale=0.85]{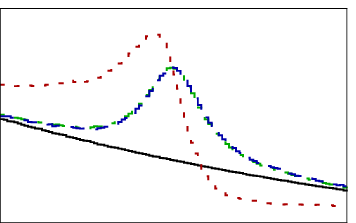} &
\includegraphics[scale=0.85]{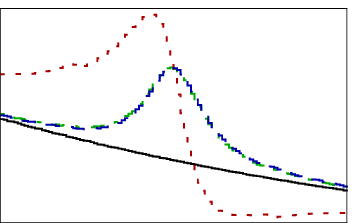} \\
\raisebox{0.35in}[0pt]{
\begin{tabular}{c}
$\phi_\ell=\frac{3\pi}{4}$ \\
$g_\ell\propto {-1+\gamma_5}$ 
\end{tabular}} &
\includegraphics[scale=0.85]{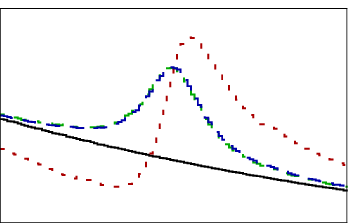} &
\includegraphics[scale=0.85]{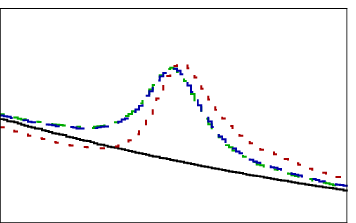} &
\includegraphics[scale=0.85]{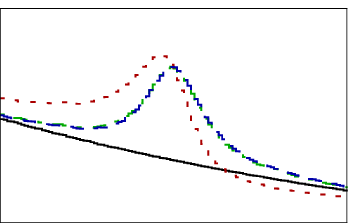} &
\includegraphics[scale=0.85]{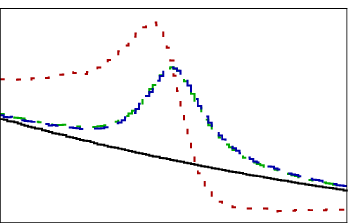} &
\includegraphics[scale=0.85]{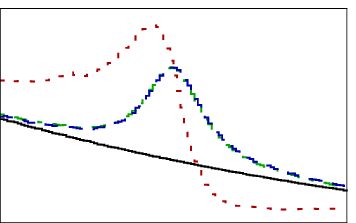} \\
\raisebox{0.35in}[0pt]{
\begin{tabular}{c}
$\phi_\ell=\frac{\pi}{2}$ \\
$g_\ell\propto\gamma_5$ 
\end{tabular}} &
\includegraphics[scale=0.85]{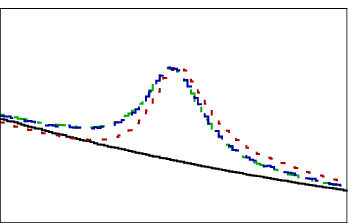} &
\includegraphics[scale=0.85]{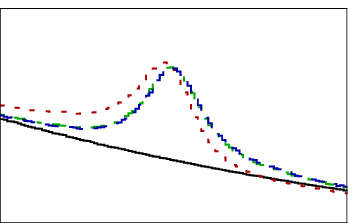} &
\includegraphics[scale=0.85]{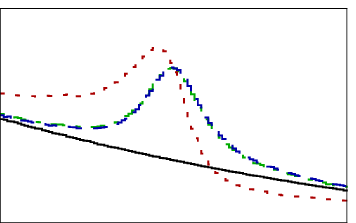} &
\includegraphics[scale=0.85]{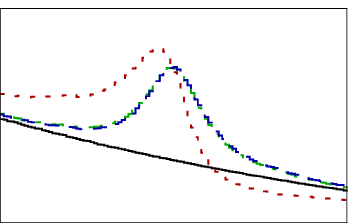} &
\includegraphics[scale=0.85]{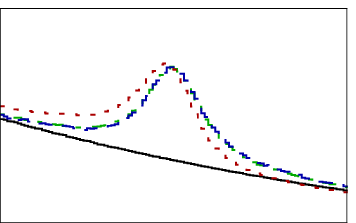} \\
\raisebox{0.35in}[0pt]{
\begin{tabular}{c}
$\phi_\ell=\frac{\pi}{4}$ \\
$g_\ell\propto {1+\gamma_5}$ 
\end{tabular}} &
\includegraphics[scale=0.85]{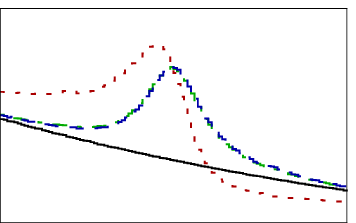} &
\includegraphics[scale=0.85]{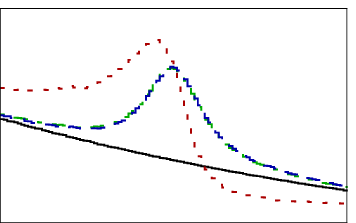} &
\includegraphics[scale=0.85]{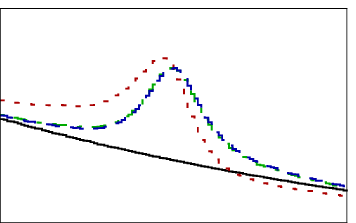} &
\includegraphics[scale=0.85]{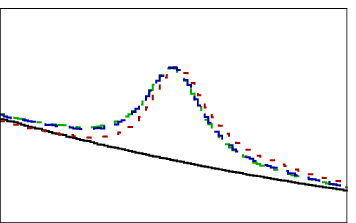} &
\includegraphics[scale=0.85]{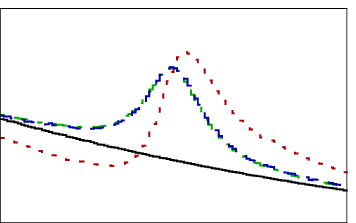} \\
\raisebox{0.35in}[0pt]{
\begin{tabular}{c}
$\phi_\ell=0$ \\
$g_\ell\propto1$ 
\end{tabular}} &
\includegraphics[scale=0.85]{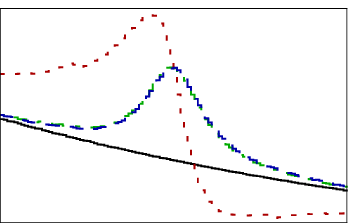} &
\includegraphics[scale=0.85]{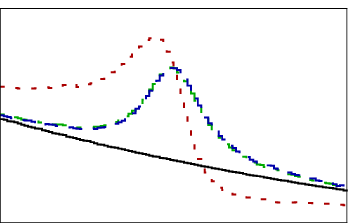} &
\includegraphics[scale=0.85]{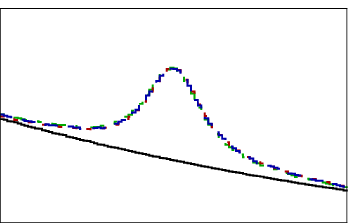} &
\includegraphics[scale=0.85]{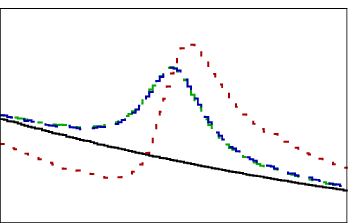} &
\includegraphics[scale=0.85]{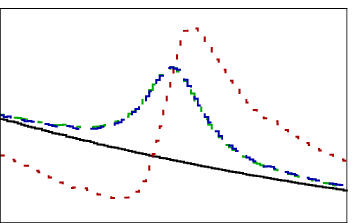} \\
&
$\phi_u=0$ &
$\phi_u=\frac{\pi}{4}$ &
$\phi_u=\frac{\pi}{2}$ &
$\phi_u=\frac{3\pi}{4}$ &
$\phi_u=\pi$ \\
&
$g_u\propto1$ &
$g_u\propto {1+\gamma_5}$ &
$g_u\propto\gamma_5$ &
$g_u\propto {-1+\gamma_5}$ &
$g_u\propto-1$ \\
\end{tabular}
\caption{
(Color online)
Invariant mass distribution for the process $pp \rightarrow\ell^+\ell^- X$.
The vertical axis is the differential cross section in arbitrary units and the horizontal axis is the dilepton invariant mass running from 950~GeV to 1050~GeV.
The row and column headers specify the nature of the chiral couplings.  
%
The solid (black) curve is for the SM, the dot-dashed (green) curve includes the scalar field, the dotted (red) curve includes the vector field, and the dashed (blue) curve includes the tensor field.  The scalar and tensor curves are indistinguishable and right on top of each other.
The mass of the new particle is taken to be 1~TeV while the width is taken to be 20~GeV.
The LHC energy is set at $7$~TeV, and the CTEQ6L PDF sets are used.
 }
 \label{fig:invmass}
\end{center}
\end{figure*}

The best observable to discover a new neutral boson coupling to quarks and charged leptons is in the spectrum of the invariant mass
\begin{equation}
M^2_{\ell \ell}=\left(p_{\ell^+}+p_{\ell^-}\right)^2 ,
\end{equation}
%
%
If there is no interference, the shape of the invariant mass distribution is of a Breit-Wigner form
and peaked at the mass of the new boson.
%
However, if there is significant  interference between the new resonance and the SM diagrams,
there can be appreciable changes in the shape.  In particular, the
peak may even be shifted.  We show the di-lepton invariant mass
distribution in Fig.~\ref{fig:invmass} for the process $pp
\rightarrow\ell^+\ell^- X$, only including the subprocess $u\bar{u}
\rightarrow\ell^+\ell^-$, mediated by a scalar (green dot-dashed), vector (red dotted) and tensor (blue dashed)
boson, along with the SM (black solid) contributions including the full spin
correlations.
To see the interference effects clearly, we have adjusted the resonance rate to be the same value as the SM background rate near the peak.
The row and column headers specify the nature of the chiral couplings.
(The scalar fields have a factor of $i$ accompanying the $\gamma_5$.) 
Here and henceforth, for illustration, the mass of the new particle is taken to be 1~TeV while the width is taken to be 20~GeV.
We adopt the parton distribution functions (PDF's) CTEQ6L \cite{Pumplin:2002vw}.
The LHC energy is set at $7$~TeV unless stated otherwise.

For
massless fermions, the scalar particle amplitude ($\mathcal{M}_S$)
is nonzero only for initial and final states of the same helicity
while the SM contribution is nonzero only for initial and final
states of opposite helicities.  For this reason, there is never any
interference between the two, and a scalar field always renders a
Breit-Wigner shape peaked at the mass of the scalar particle.  This
can be seen analytically in Table~\ref{tab:helam_neutral}, and also numerically
in the dot-dashed (green) curves of Fig.~\ref{fig:invmass}.
%
%

The new vector boson does interfere with the SM.  The amount of interference depends on the parity properties of the couplings.  The interference always flips sign at the mass of the new boson due to the phase change in the propagator $s-M_V^2$.  In addition, there is an overall sign coming from the couplings. 
%
After summing over helicities and integrating over the scattering angle $\theta$, we find that the interference with the SM photon diagram is given by
\begin{equation}
\sum_{hel}\int_{-1}^1d\cos\theta\left(\mathcal{M}_\gamma\mathcal{M}_V^*+\mathcal{M}_\gamma^*\mathcal{M}_V\right)
\propto\mathcal{R}\left(g_{Vmi}^qg_{Vjn}^\ell\right) ~,
\end{equation}
where $\mathcal{R}(x)$ means the real part of $x$.  The interference with the photon is independent of parity violation and only depends on the sign of the vectorial coupling. The interference with the $Z$ boson diagram, on the other hand, is given by
\begin{eqnarray}
&& \sum_{hel}\int_{-1}^1d\cos\theta\left(\mathcal{M}_Z\mathcal{M}_V^*+\mathcal{M}_Z^*\mathcal{M}_V\right)\propto\\
&& \mathcal{R}\left[\left(g^q_{ZV} g_{Vmi}^q + g^q_{ZA}
g_{Ami}^q\right)\left(g^\ell_{ZV} g_{Vjn}^\ell + g_{ZA}^\ell
g_{Ajn}^\ell\right)\right] , \nonumber
\end{eqnarray}
which has more complicated dependence on the vector and axial vector
couplings.  Since $g_{ZV}^\ell$ is very small, the sign of the
interference is more strongly dependent on the axial coupling to leptons.  
Again, the interference can be seen analytically in Table~\ref{tab:helam_neutral} and numerically in the dotted (red) curves of Fig.~\ref{fig:invmass}.
%

In the case of the tensor field, the term proportional to $d^2_{m,m'}$ does not contribute to interference in the invariant mass distribution since $d^2_{m,m'}$ is orthogonal to $d^1_{m,m'}$ and this interference vanishes after integration over $\theta$.  The $d^1_{m,m'}$ term of the tensor amplitude is due to the off-shell effects and
does not contribute to the peak at $M_{\ell\ell}=M_V$. Consequently, the interference is very weak and not readily observable
 in the continuum invariant mass distribution.
The final result is that the tensor boson has a Breit-Wigner shape
peaked at its mass as can be seen in the dashed (blue) curves of
Fig.~\ref{fig:invmass}, which is hardly distinguishable from the
case of a scalar.

\begin{figure*}[tb]
\begin{center}
\begin{tabular}{cccccc}
\raisebox{0.35in}[0pt]{
\begin{tabular}{c}
$\phi_\ell=\pi$ \\
$g_\ell\propto-1$ 
\end{tabular}} &
\includegraphics[scale=0.85]{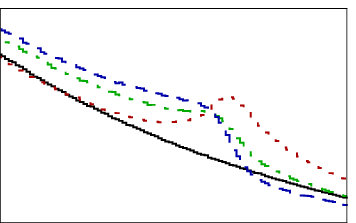} &
\includegraphics[scale=0.85]{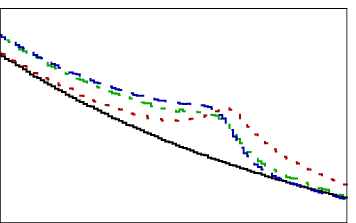} &
\includegraphics[scale=0.85]{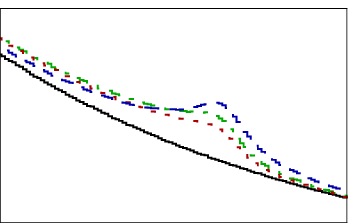} &
\includegraphics[scale=0.85]{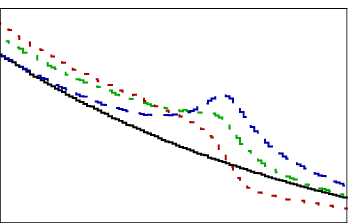} &
\includegraphics[scale=0.85]{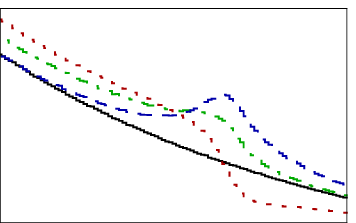} \\
\raisebox{0.35in}[0pt]{
\begin{tabular}{c}
$\phi_\ell=\frac{3\pi}{4}$ \\
$g_\ell\propto {-1+\gamma_5}$ 
\end{tabular}} &
\includegraphics[scale=0.85]{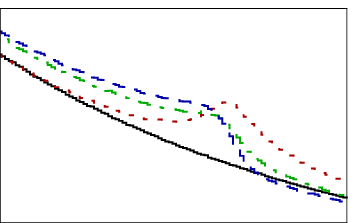} &
\includegraphics[scale=0.85]{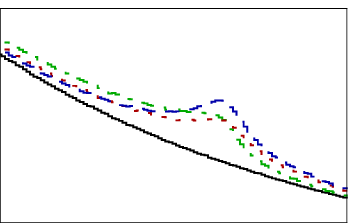} &
\includegraphics[scale=0.85]{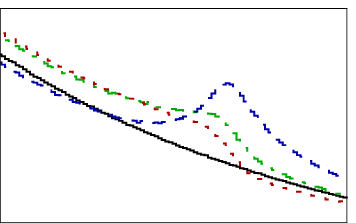} &
\includegraphics[scale=0.85]{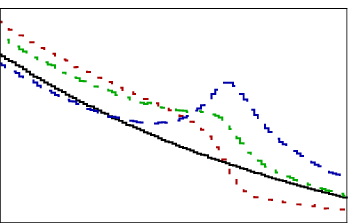} &
\includegraphics[scale=0.85]{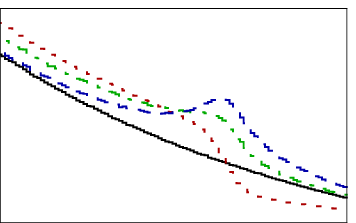} \\
\raisebox{0.35in}[0pt]{
\begin{tabular}{c}
$\phi_\ell=\frac{\pi}{2}$ \\
$g_\ell\propto\gamma_5$ 
\end{tabular}} &
\includegraphics[scale=0.85]{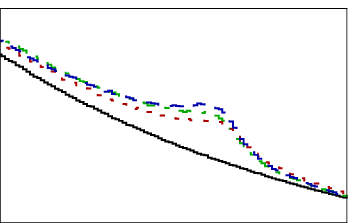} &
\includegraphics[scale=0.85]{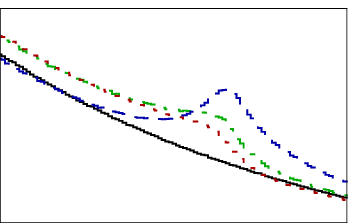} &
\includegraphics[scale=0.85]{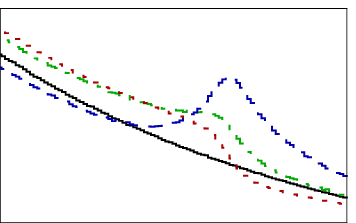} &
\includegraphics[scale=0.85]{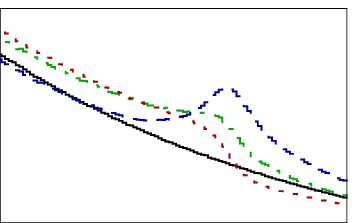} &
\includegraphics[scale=0.85]{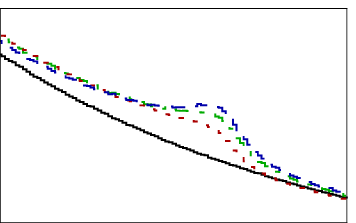} \\
\raisebox{0.35in}[0pt]{
\begin{tabular}{c}
$\phi_\ell=\frac{\pi}{4}$ \\
$g_\ell\propto {1+\gamma_5}$ 
\end{tabular}} &
\includegraphics[scale=0.85]{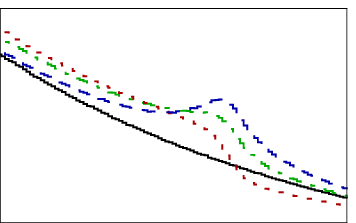} &
\includegraphics[scale=0.85]{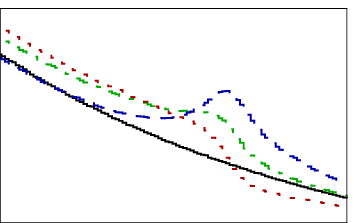} &
\includegraphics[scale=0.85]{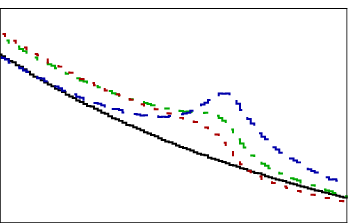} &
\includegraphics[scale=0.85]{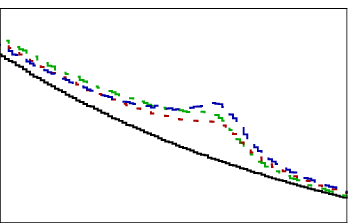} &
\includegraphics[scale=0.85]{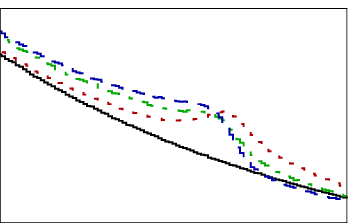} \\
\raisebox{0.35in}[0pt]{
\begin{tabular}{c}
$\phi_\ell=0$ \\
$g_\ell\propto1$ 
\end{tabular}} &
\includegraphics[scale=0.85]{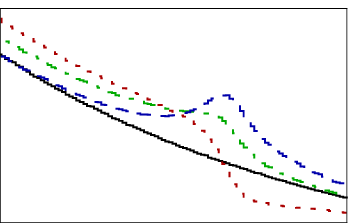} &
\includegraphics[scale=0.85]{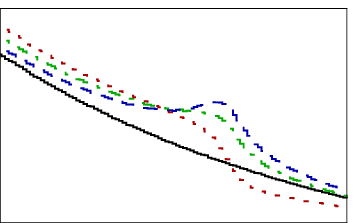} &
\includegraphics[scale=0.85]{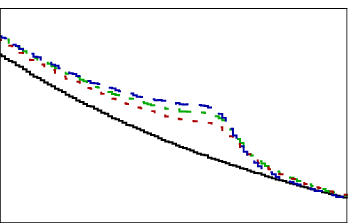} &
\includegraphics[scale=0.85]{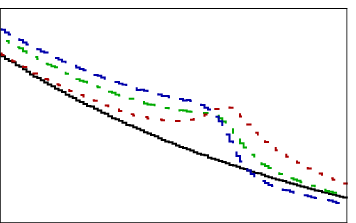} &
\includegraphics[scale=0.85]{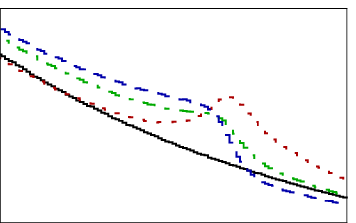} \\
&
$\phi_u=0$ &
$\phi_u=\frac{\pi}{4}$ &
$\phi_u=\frac{\pi}{2}$ &
$\phi_u=\frac{3\pi}{4}$ &
$\phi_u=\pi$ \\
&
$g_u\propto1$ &
$g_u\propto {1+\gamma_5}$ &
$g_u\propto\gamma_5$ &
$g_u\propto {-1+\gamma_5}$ &
$g_u\propto-1$ \\
\end{tabular}
\caption{
(Color online)
Transverse momentum distribution of $\ell^-$ for the process $pp \rightarrow\ell^+\ell^- X$.
The vertical axis is the differential cross section in arbitrary units and the horizontal axis is the transverse momentum running from 450~GeV to 525~GeV. The row and column headers specify the nature of the chiral couplings.
%
The curve legends are the same as in Fig.~\ref{fig:invmass}.
 }
 \label{fig:pT}
\end{center}
\end{figure*}

\subsection{Transverse momentum distribution\label{sec:pT}  } 

Once  a new boson resonance is established in the invariant mass spectrum, it will be of ultimate importance to study its other quantum numbers, such as its spin, chiral couplings, etc. 
%
For a massless particle, the transverse momentum is related to the scattering angle
\begin{equation}
{p_T} = E \sin\theta .
\end{equation}
Thus, the differential distribution of $p_{T}$ may contain additional information of chiral interactions of the resonance via the interference
with the SM diagrams.
%
We present the transverse momentum distribution of the negatively charged lepton for a variety of parity violation cases in Fig.~\ref{fig:pT}. 
As expected, there is no interference between a spin-0 state and the SM diagrams. 
The fact that a spin-2 resonance interferes with the SM in the transverse momentum distribution gives us a potential new way of determining the spin of the resonance, unlike the invariant mass distribution.
If we find interference present in both the invariant mass distribution and the transverse momentum distribution, we can conclude that the new resonance is a spin-1 particle.  
Furthermore, 
the amount of interference and the sign of the interference can give information about the size and sign of the parity violation in the couplings.
We emphasize that the analysis of the $p_T$ distribution does not require a
knowledge of the quark moving direction, nor the reconstruction of the CM
frame.

\subsection{Angular distribution
\label{sec:v-angledist}}

One of the main advantages of the present process is the feasibility to fully reconstruct the CM system of the two charged leptons that is the rest frame of the new boson.  Although we do not know the direction of the quark on an event-by-event basis, it is strongly correlated with the direction of the CM frame of the charged lepton pair due to the parton distribution functions of the quark versus the anti-quark in a proton \cite{Langacker:1984dc}.
We calculate this angle by first boosting into the CM frame of the charged leptons, and then taking the angle between the moving direction of the negatively charged lepton and the direction of the boost.  In the case of the gluon initial state, both directions are equally valid and we simply use the direction of the boost to measure the negatively charged lepton as in the quark case.

\begin{figure*}[tb]
\begin{center}
\includegraphics[scale=1.05,bb= 0 0 2.4in 1.7in,clip=true]{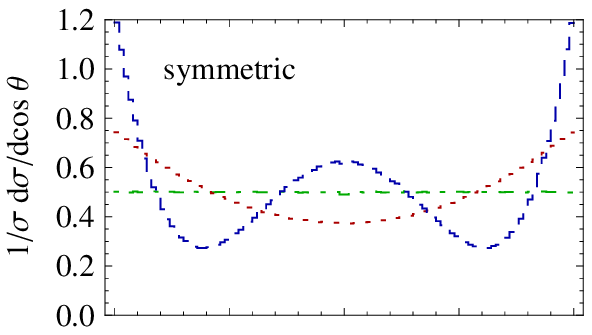}
\hspace{-0.35in}
\raisebox{0.015in}[0pt]{\includegraphics[scale=1,bb= 0 0 2.4in 1.7in,clip=true]{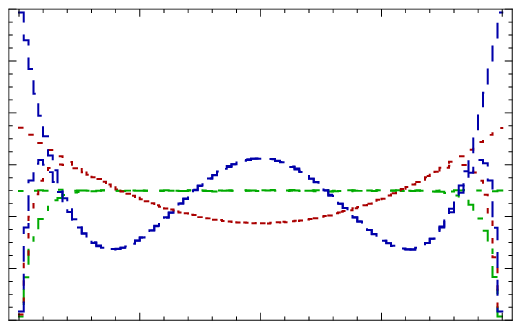}}\\
\vspace{-0.4in}
\includegraphics[scale=1.05,bb= 0 0 2.4in 1.7in,clip=true]{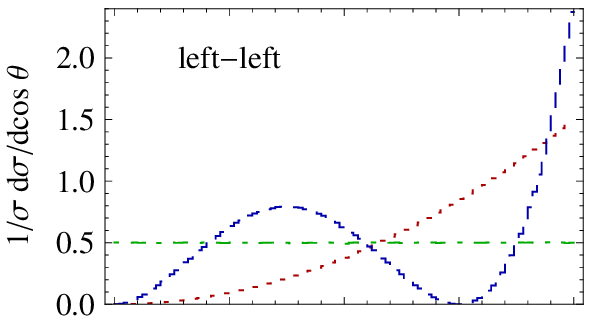} 
\hspace{-0.35in}
\raisebox{0.015in}[0pt]{\includegraphics[scale=1,bb= 0 0 2.4in 1.7in,clip=true]{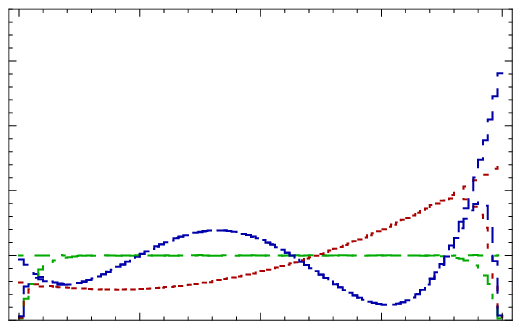}}\\
\vspace{-0.4in}
\includegraphics[scale=1.05,bb= 0 0 2.4in 1.7in,clip=true]{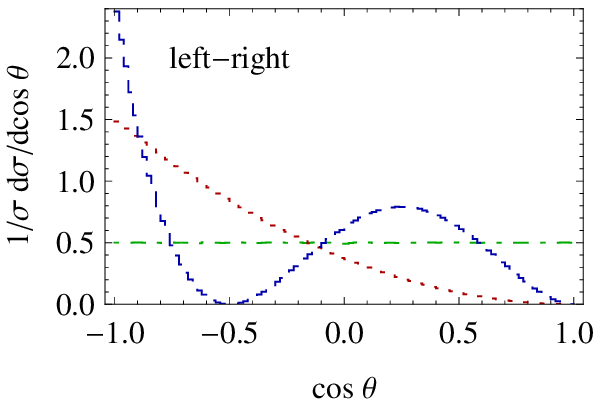} 
\hspace{-0.35in}
\raisebox{0.015in}[0pt]{\includegraphics[scale=1,bb= 0 0 2.4in 1.7in,clip=true]{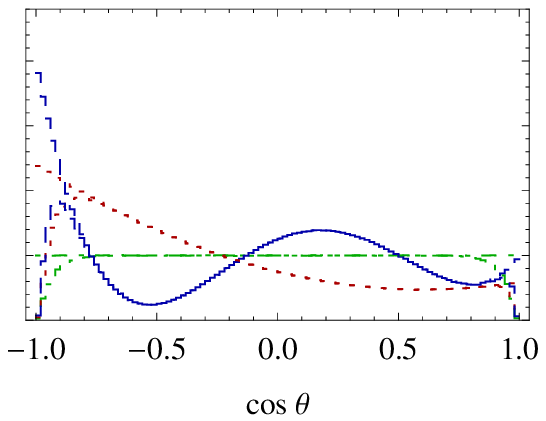}}
\caption{
(Color online)
Angular distributions  of $\ell^{-}$  in the CM frame
for the process $pp \rightarrow\ell^+\ell^- X$. The curve legends are the same as in Fig.~\ref{fig:invmass}.
In the top row purely scalar, vectorial and tensorial couplings are used.  In the second row left chiral couplings are used while in the third row left chiral couplings are used for the quark but right chiral couplings are used for the leptons.  In the left column the PDF's are turned off and the lab frame is the same as the CM frame, while on the right, the PDF's are turned on and the boost direction is used.  
On the right panels, the visible bend-down feature in the forward-backward regions 
is due to the kinematical cuts  as described in the text.
\label{fig:cos theta neutral grid} }
\end{center}
\end{figure*}

The angular dependence comes from the Wigner $d^j_{m,m'}$ functions.  For
each helicity combination and each spin for the new boson, these are
determined by the kinematics (see Appendix~\ref{app:d functions}).
The mixture of the $d^j_{m,m'}$ functions encodes the information of
its spin and chiral interactions.  For the scalar field, only
$d^0_{0,0}$ contributes and so the angular distribution is flat.
This can be seen in the dot-dashed (green) curves of
Fig.~\ref{fig:cos theta neutral grid}.  For a vector field, the
$d^1_{\pm1,\pm1}$ functions contribute.  Each one is squared and
summed with the appropriate factors (see
Table~\ref{tab:helam_neutral}). This gives the angular distribution
\begin{eqnarray}
\overline{\sum} |{\cal M}_{V}|^{2} &=& \frac{4s^2}{\left| D_{V}\right|^2} \left[
A_V\left(1+\cos\theta\right)^2+B_V\left(1-\cos\theta\right)^2\right] ,
\nonumber \\
\label{angle}
A_V&=&
\left(g_{_{VR}}^q  g_{_{VR}}^\ell\right)^2+\left(g_{_{VL}}^q g_{_{VL}}^\ell\right)^2,  \\
B_V &=&
\left(g_{_{VR}}^q  g_{_{VL}}^\ell\right)^2+\left(g_{_{VL}}^q  g_{_{VR}}^\ell\right)^2 .\nonumber
\end{eqnarray}
This
is the formula of a parabola versus $\cos\theta$ where the amount of
parity violation determines where the minimum lies.  This parabolic
shape can be seen in the dotted (red) curves of Fig.~\ref{fig:cos theta neutral grid}.
We find that the angular
distribution is symmetric whenever either coupling is pure vector or
axial vector.
%
If both the quark and the lepton couplings are parity-violating (mixtures of 1 and $\gamma_5$ terms), then the angular distribution is shifted to one side or the other.  The shift is maximal when both couplings are purely chiral (the magnitude of the vector coupling equals that of the axial coupling,
as in $1\pm \gamma_{5}$). However, it is interesting to note that the distributions are identical for either purely right-handed or purely left-handed, and are parity-transformed for right- and left-mixed couplings.

For the tensor field, the angular dependence is mainly a mixture of the $d^2_{\pm1,\pm1}$ functions (the $d^1_{\pm1,\pm1}$ functions give a small perturbation)
\begin{eqnarray}
\overline{\sum} |{\cal M}_{T}|^{2} &=& \frac{16s^4}{\Lambda^4 \left| D_{T} \right|^2 }
\left[ A_T\left(1+\cos\theta\right)^2\left(2\cos\theta-1\right)^2 \right. \nonumber\\
&& \qquad + \left. B_T\left(1-\cos\theta\right)^2\left(2\cos\theta+1\right)^2 \right] , \nonumber
\end{eqnarray}
where $A_{T}$ and $B_{T}$ are of the same form as in Eq.~(\ref{angle}),
but with the tensor couplings. This is a quartic function with one
local maximum and two local minima, a ``W'' shape.  It can be seen
in the dashed (blue) curves of Fig.~\ref{fig:cos theta neutral grid}.
In particular, we find that
if either of the couplings are purely tensor ($g\propto1$) or purely
axial tensor ($g\propto\gamma_5$) the distribution is symmetric,
with the local maximum occurring at $\cos\theta=0$. However, if both
quark and lepton couplings are mixtures of $1$ and $\gamma_5$, then
the curves shift toward one side or the other, leading to parity
violation.  As in the vector case, the shift is maximal when the
coefficient of $1$ has the same magnitude as that of $\gamma_5$.

To be more realistic, we fold in the CTEQ6L parton distribution functions and adopt some basic acceptance cuts on the
transverse momentum and its pseudo-rapidity 
for the charged leptons
\begin{equation}
p_{T\ell} >20\ {\rm GeV}, \quad  | \eta_{\ell} | < 2.5.
\label{eq:llcut}
\end{equation}
We see the effects in the three panels on the right-hand side in Fig.~\ref{fig:cos theta neutral grid}.
Due to the misidentification of the correct quark momentum direction, the far forward/backward regions are diluted. 
The rapidity cut limits the angle reach in the same region as seen at the drop near 0.8.

%

\subsection{Forward-backward asymmetry
\label{sec:v-fba}}

It is customary to construct the forward-backward asymmetry, based on the partially integrated rates in the opposite
angular regions,
which can be defined as
\begin{eqnarray}
A_{FB}(c_0) = \frac{N(\cos\theta > c_0) - N(\cos\theta < -c_0)}
{N(\cos\theta > c_0) + N(\cos\theta < -c_0)} ~.
\end{eqnarray}
$c_0=0$ will lead to the largest event rate, while a particular choice of $c_{0}$ may optimize the
size of the asymmetry\footnote{The asymmetry can be defined with any differential form with respect to the angular range.  Other Lorentz invariant asymmetries have also been proposed, for example \cite{Ferrario:2008wm,Hewett:2011wz}.  Others have noted that the asymmetry can be used to discriminate between different $Z'$ models \cite{Dittmar:1996my,Zhou:2011dg,Godfrey:1987qz,Rosner:1986cv,Rosner:1995ft,delAguila:1993ym,Wang:2010tg,Barger:1986hd,Osland:2009tn,Diener:2009vq}.}.

It is clear that a scalar field will give $A_{FB}=0$ for any value of $c_0$ since it has a flat angular distribution.

For a vector field, it is natural to use $c_0=0$.  It can be shown that the forward-backward asymmetry at the peak of
a vector resonance is given by
\begin{eqnarray}
&& A_{FB}^V = A_{FB}(0) = \frac{3}{4}\frac{A_V-B_V}{A_V+B_V} \label{eq:FBA1} \\
&& =  \frac{3 g_V^q g_V^\ell g_A^q g_A^\ell}{\left[ (g_V^q)^2 + (g_A^q)^2 \right] \left[ (g_V^\ell)^2 + (g_A^\ell)^2 \right] }
= \frac{3}{4} \sin2\phi_q \sin2\phi_\ell , \nonumber
\end{eqnarray}
where $\phi_f$ defines the ratio of vector and axial couplings and is given by
\begin{eqnarray}
\nonumber
\cos\phi_f = \frac{g_V^f}{\sqrt{\left(g_V^f\right)^2+\left(g_A^f\right)^2}},\
\sin\phi_f = \frac{g_A^f}{\sqrt{\left(g_V^f\right)^2+\left(g_A^f\right)^2}}.
\end{eqnarray}
For example, $\phi_f=0$ corresponds to a pure vector coupling with $g^f\propto1$, $\phi_f=\pi/4$ corresponds to a pure chiral coupling with $g^f\propto1+ \gamma_5$, and so on.  Eq.~(\ref{eq:FBA1}) shows that the absolute value of the asymmetry is bounded to be less than or equal to 3/4.  Also, if any of the couplings $g_V^q$, $g_A^q$, $g_V^\ell$ and $g_A^\ell$ is identically zero, there is no forward-backward asymmetry, resulting in parity conservation.
The SM asymmetry for the $u\bar{u}\rightarrow \ell^+\ell^-$ process  at $\sqrt{s}\gg M_Z$ is approximately $0.6$.
As mentioned before, the quark momentum direction is ambiguous in $pp$ collisions. We thus
take the angle in the CM frame with respect to the boost direction,
which is more likely to be the direction of the quark. 

\begin{figure*}[!tb]
\begin{center}
\begin{tabular}{cc}
\multicolumn{2}{c}{Vector}\\
\includegraphics[scale=1.2]{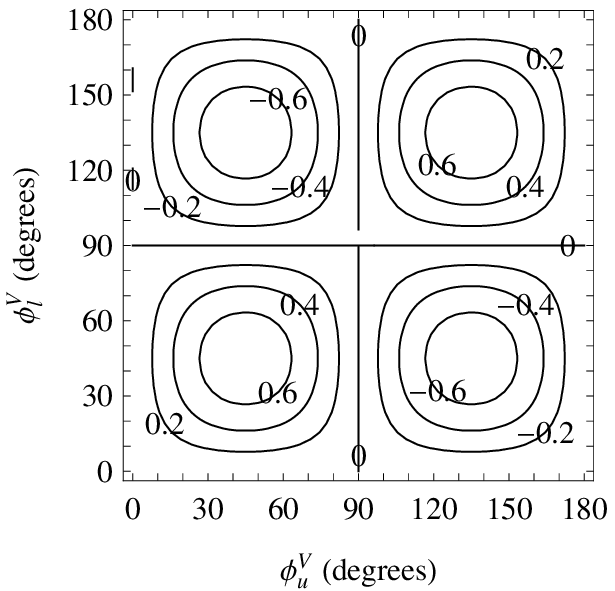} &
\includegraphics[scale=1.2]{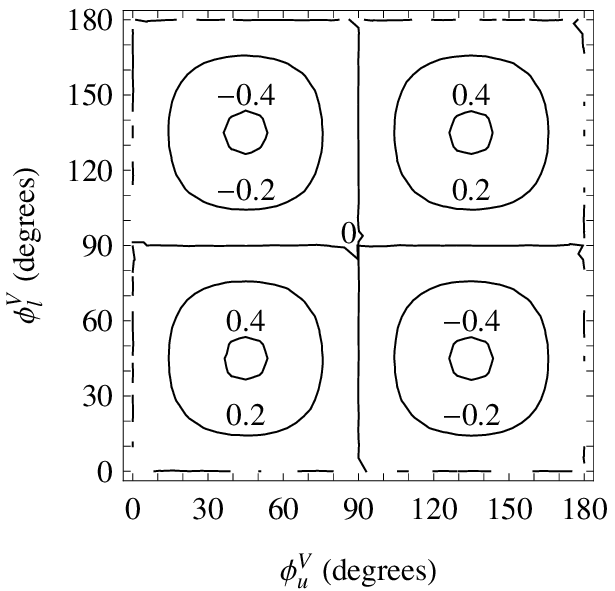} \\
(a) & (b) \\\\
\multicolumn{2}{c}{Tensor}\\
\includegraphics[scale=1.2]{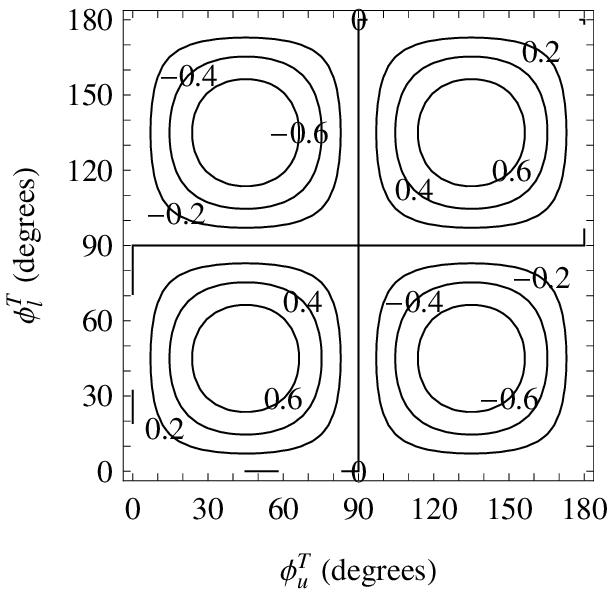} &
\includegraphics[scale=1.2]{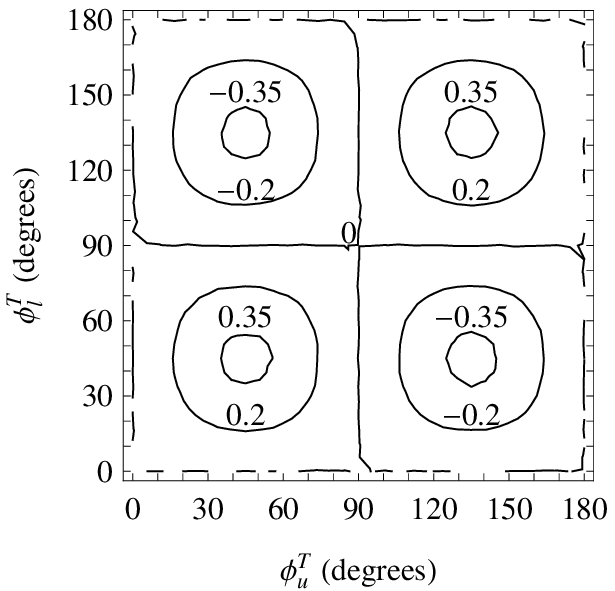}\\
(c) & (d)
\end{tabular}
\end{center}
\caption{Contours of forward-backward asymmetry labelled by  the values of $A_{FB}$
in the $u\bar{u}\rightarrow \ell^+\ell^-$ process. In the first row (a) and (b),  a vector resonance is presented and in the second row (c) and (d), a tensor resonance is presented. 
Plots (a) and (c) show the result with the new physics amplitude alone in the lab frame while plots (b) and (d) include the PDF's and acceptance cuts.
\label{fig:FBA1}}
\end{figure*}

Again, to be more realistic, we convolute with the parton distribution functions, and take the acceptance cuts as in Eq.~(\ref{eq:llcut}).
In Fig.~\ref{fig:FBA1}(a) and \ref{fig:FBA1}(b), we present the equal-valued
contours of forward-backward asymmetry for a vector resonance in the
$\phi_u^{V}$-$\phi_\ell^{V}$ plane.  Fig. \ref{fig:FBA1}(a) shows
contours for  $u\bar{u}\rightarrow \ell^+\ell^-$ via the resonance
only, and Fig. \ref{fig:FBA1}(b) includes the PDF convolution at the
hadronic level.
We see that the asymmetry is reduced when the parton distribution functions are turned on and acceptance cuts imposed.  
The SM value becomes $\sim0.37$. As for the signal, the maximum is reduced from $3/4$ to $\sim0.41$.  The reason for this is that the boost direction corresponds with the quark direction much of the time but is sometimes in the opposite direction. The asymmetry breaks up into quadrants separated by the angles $\phi^V=n\pi/2$ for some integer $n$ where the asymmetry is $0$.  This is where the sign of the vectorial or axial coupling changes sign, thus changing the sign of the asymmetry. 
%
%

For a tensor particle, we find that $c_0\neq0$ is required to obtain a non-zero asymmetry.
This can be seen by noting that for $c_0=0$,
\begin{eqnarray}
\int_{-1}^0d\cos\theta\Big[&
A_T\left(1+\cos\theta\right)^2\left(2\cos\theta-1\right)^2&\nonumber\\
&+B_T\left(1-\cos\theta\right)^2\left(2\cos\theta+1\right)^2&\Big]\nonumber\\
=\int_0^1d\cos\theta\Big[&
A_T\left(1+\cos\theta\right)^2\left(2\cos\theta-1\right)^2&\nonumber\\
&+B_T\left(1-\cos\theta\right)^2\left(2\cos\theta+1\right)^2&\Big]\nonumber
\end{eqnarray}
for all values of $A_T$ and $B_T$.  In other words, although the tensor distribution is asymmetric, it always shifts in such a way as to have equal area under the angular distribution for positive and negative values of $\cos\theta$.
For this reason, it is necessary to use $c_0\neq0$.  We find that the difference is maximized when $c_0=1/\sqrt{2}$.
\begin{eqnarray}
\label{equation:ATFB}
A^T_{FB} &=& A_{FB}\left(\frac{1}{\sqrt{2}}\right) \nonumber\\
&=& \frac{5}{16-7\sqrt{2}}\frac{A_T-B_T}{A_T+B_T}\nonumber\\
&=& \frac{20}{16-7\sqrt{2}}\frac{g_T^q g_T^\ell g_{AT}^q g_{AT}^\ell}
{\left[ (g_T^q)^2 + (g_{AT}^q)^2 \right] \left[ (g_T^\ell)^2 + (g_{AT}^\ell)^2 \right]}\nonumber\\
&=& \frac{5}{16-7\sqrt{2}}\sin2\phi_q^T\sin2\phi_\ell^T ,
\end{eqnarray}
where $\phi_f^T$ defines the ratio of tensor and axial tensor couplings, similar to the vector case.
%
Similar to Figs.~\ref{fig:FBA1}(a) and \ref{fig:FBA1}(b),
the equal-valued contours of forward-backward asymmetry for a tensor resonance are presented
in Figs~\ref{fig:FBA1}(c) and \ref{fig:FBA1}(d). We again see that the asymmetry is reduced when the parton distribution functions are turned on.  The maximum asymmetry is reduced to a little greater than $\sim0.37$.  The sign is similar to the vector case.

If an experiment discovers a resonance in the invariant mass distribution of the 
neutral Drell-Yan process, measuring the forward-backward asymmetry
could help distinguish the spins in the early days before enough
events are accumulated for reconstructing the angular distribution.
If a significant asymmetry is found in $A_{FB}(0)$, this would be
evidence of a new vector resonance.  If no asymmetry is found in
$A_{FB}(0)$, then one should further determine $A_{FB}(1/\sqrt{2})$.
If $A_{FB}(1/\sqrt{2})$ is significantly different from zero while
$A_{FB}(0)=0$, then this would be evidence of a new tensor particle.
If both are zero, then this is evidence for a symmetric
distribution, but could come from any of the spins we have discussed
here.

%
%

\section{Charged Boson Resonance
\label{sec:vprime}}

\begin{figure*}[tb]
\begin{center}
\begin{tabular}{p{0.8in}p{0.8in}p{0.8in}p{0.8in}p{0.8in}p{0.8in}p{0.8in}p{0.8in}}
\multicolumn{2}{c}{\includegraphics[scale=.43]{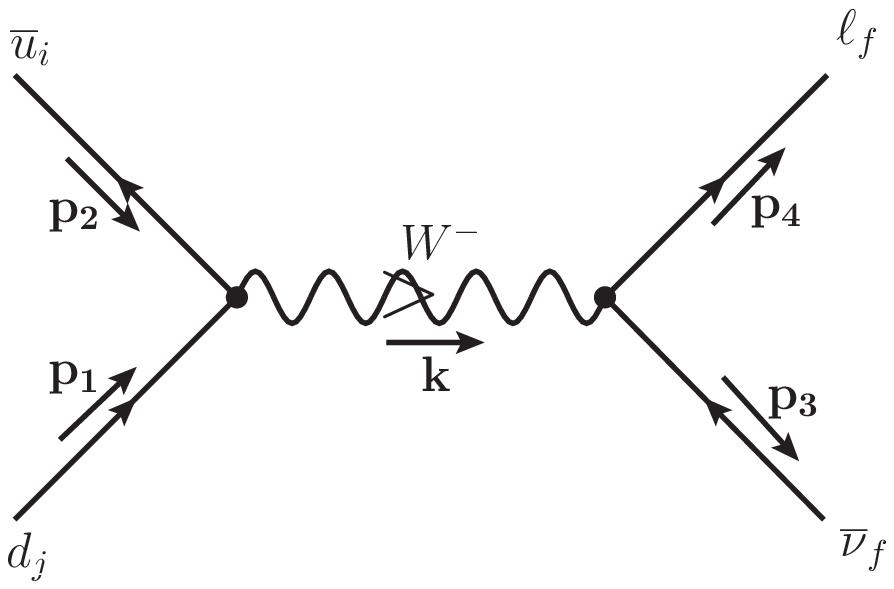}} &
\multicolumn{2}{c}{\includegraphics[scale=.43]{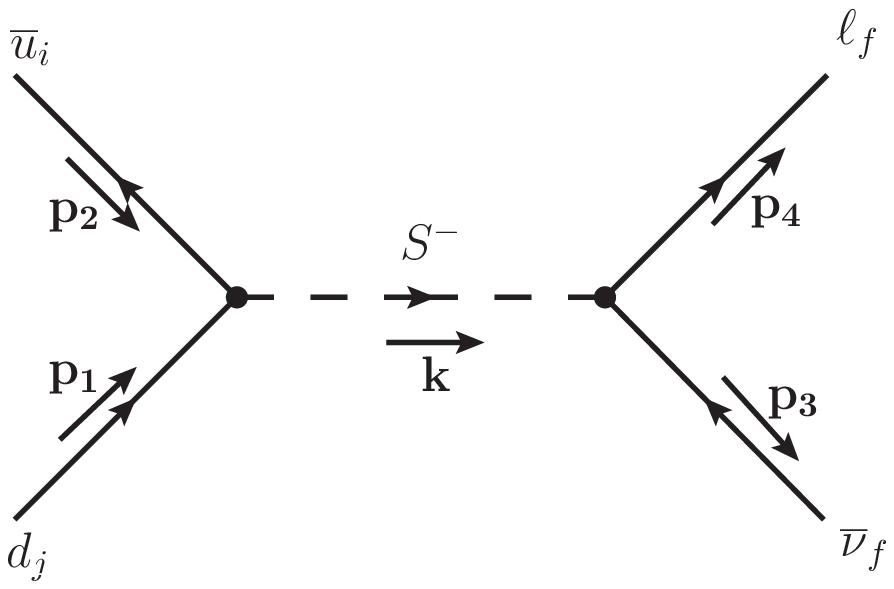}} &
\multicolumn{2}{c}{\includegraphics[scale=.43]{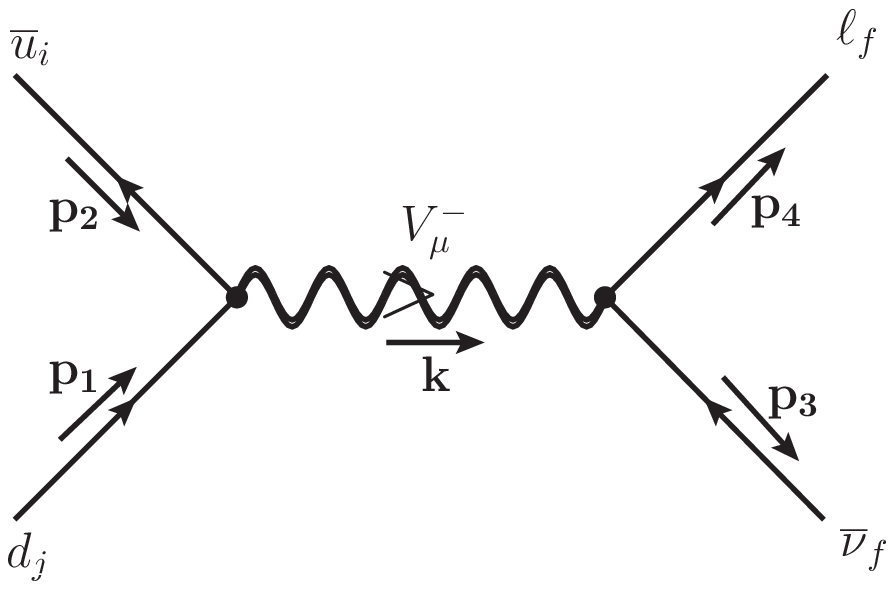}} &
\multicolumn{2}{c}{\includegraphics[scale=.43]{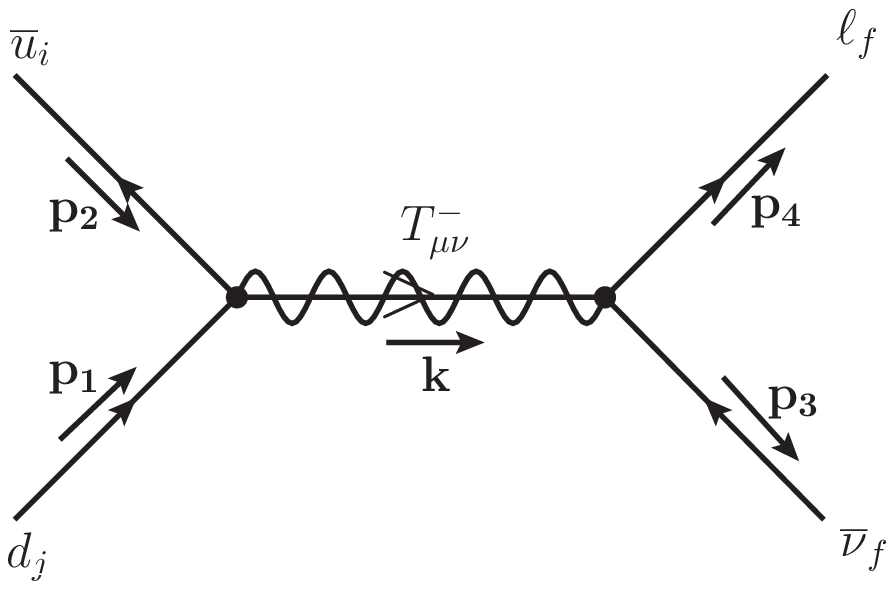}}\\
\multicolumn{2}{c}{($W$)} &
\multicolumn{2}{c}{($S$)} &
\multicolumn{2}{c}{($V$)} &
\multicolumn{2}{c}{($T$)} \end{tabular}
\caption{\label{scatter_diagram_W} 
The $s$-channel Feynman diagrams responsible for $pp\rightarrow\ell\bar{\nu}X$.  Contributions to this process are from a $W$ boson ($W$), a new scalar particle ($S$), a new vector particle ($V$) and a new tensor particle ($T$).}
\end{center}
\end{figure*}

Charged bosons contribute to the Drell-Yan process
$pp\rightarrow\ell^{\pm} \nu X$ through the diagrams in
Fig.~\ref{scatter_diagram_W} at tree level.  The resulting
amplitudes are shown in Table~\ref{tab:helam_charged}.

\begin{table*}[tb]
\begin{center}
\renewcommand{\arraystretch}{2.5}
\renewcommand{\tabcolsep}{0.25in}
\begin{tabular}{|ll|}

\multicolumn{2}{c} {\boldmath $d_{i}(\lambda) \bar u_{m}(-\lambda)\rightarrow \ell_{j}(\lambda') \bar \nu_{n}(-\lambda')$ }  \\  \hline\hline
$\mathcal{M}^{\lambda\lambda'}_W=\mathcal{C}^{\lambda\lambda'}_W\delta_{\lambda',-1}d^{1}_{1,1}$ & $\mathcal{C}^{\lambda\lambda'}_W=-\frac{g^2_WV^{KM}_{ij}s}{D_W}\delta_{\lambda,-1}$\\

$\mathcal{M}^{\lambda\lambda'}_{V}=\mathcal{C}^{\lambda\lambda'}_V\,d^{1}_{1,\lambda\lambda'}$ & $\mathcal{C}^{\lambda\lambda'}_V=-\frac{2s}{D_{V^\pm}}(\lambda h^q_V+h^q_A)_{mi}(\lambda' h^{\ell\dagger}_V+h^{\ell\dagger}_A)_{jn}$\\

$\mathcal{M}^{\lambda\lambda'}_{T}=\mathcal{C}^{\lambda\lambda'}_{T2}\,d^{2}_{1,\lambda\lambda'}+\mathcal{C}^{\lambda\lambda'}_{T1}\,d^{1}_{1,\lambda\lambda'}$ & $\mathcal{C}^{\lambda\lambda'}_{T2}=-\frac{2\lambda\lambda's^2}{\Lambda^2 D_{T^\pm}}\mathcal{T}_+(h^{q}_T+\lambda h^{q}_{AT})_{mi}\mathcal{T}_+(h^{\ell\dagger}_T+\lambda'h^{\ell\dagger}_{AT})_{jn}$\\

& $\mathcal{C}^{\lambda\lambda'}_{T1}=-\frac{2\lambda\lambda's^2(s-M^2_{T^\pm})}{\Lambda^2M^2_{T^\pm}D_{T^\pm}}\mathcal{T}_-(h^{q}_T+\lambda h^{q}_{AT})_{mi}\mathcal{T}_-(h^{\ell\dagger}_T+\lambda'h^{\ell\dagger}_{AT})_{jn}$\\

\hline

\multicolumn{2}{c}{\boldmath $d_{i}(\lambda) \bar u_{m}(\lambda)\rightarrow \ell_{j}(\lambda') \bar \nu_{n}(\lambda')$ } \\\hline\hline

$\mathcal{M}^{\lambda\lambda'}_S=\mathcal{C}^{\lambda\lambda'}_S\,d^{0}_{0,0}$ & $\mathcal{C}^{\lambda\lambda'}_S=\frac{s}{D_{S^\pm}}(i\lambda h^q_S-h^{q}_P)_{mi}(i\lambda'h^{\ell\dagger}_S+h^{\ell\dagger}_P)_{jn}$ \\

\hline
\end{tabular}
\caption{\label{tab:helam_charged}Helicity scattering amplitudes for
the parton-level processes.  The amplitudes correspond with the
diagrams in Fig.~\ref{scatter_diagram_W}.  The particles in the $s$-channel exchange are labeled by subscripts ($W$, $V$, $T$ and $S$).
$\lambda$ and $\lambda'$ are the helicities and we define
$\mathcal{T}_\pm(M)=M\pm\tilde{M}$ where $\tilde{M}$ means that we
replace $h$ by $\tilde{h}$ in $M$. }
\end{center}
\end{table*}

\subsection{Transverse mass distribution
\label{sec:vprime-transmass}}

\begin{figure*}[tb]
\begin{center}
\begin{tabular}{cccccc}
\raisebox{0.35in}[0pt]{
\begin{tabular}{c}
$\phi_\ell=\pi$ \\
$g_\ell\propto-1$ 
\end{tabular}}  &
\includegraphics[scale=0.85]{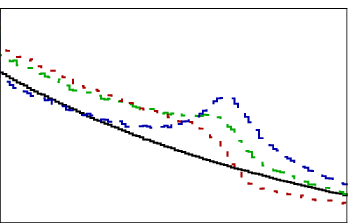} &
\includegraphics[scale=0.85]{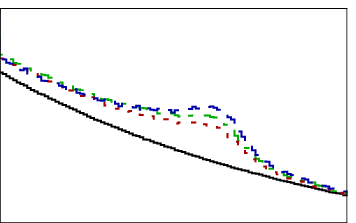} &
\includegraphics[scale=0.85]{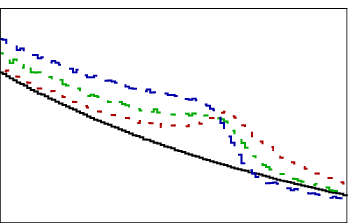} &
\includegraphics[scale=0.85]{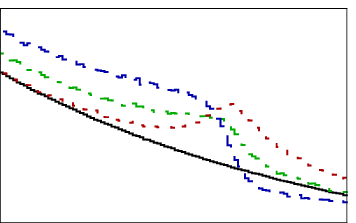} &
\includegraphics[scale=0.85]{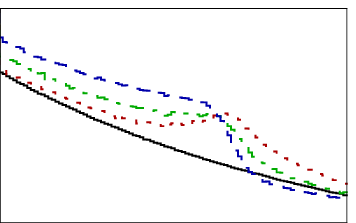} \\
\raisebox{0.35in}[0pt]{
\begin{tabular}{c}
$\phi_\ell=\frac{3\pi}{4}$ \\
$g_\ell\propto {-1+\gamma_5} $ 
\end{tabular}} &
\includegraphics[scale=0.85]{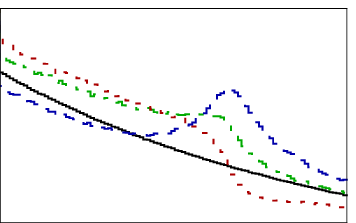} &
\includegraphics[scale=0.85]{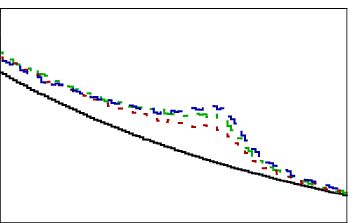} &
\includegraphics[scale=0.85]{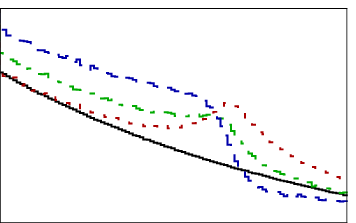} &
\includegraphics[scale=0.85]{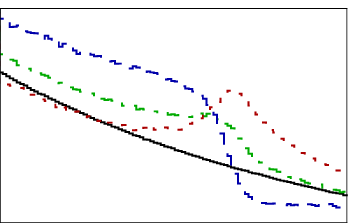} &
\includegraphics[scale=0.85]{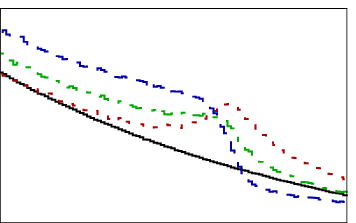} \\
\raisebox{0.35in}[0pt]{
\begin{tabular}{c}
$\phi_\ell=\frac{\pi}{2}$ \\
$g_\ell\propto\gamma_5$ 
\end{tabular}} &
\includegraphics[scale=0.85]{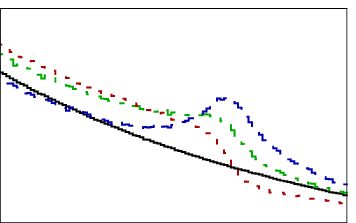} &
\includegraphics[scale=0.85]{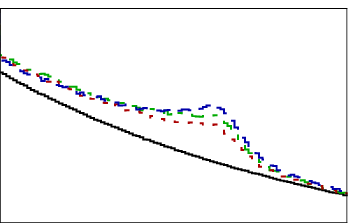} &
\includegraphics[scale=0.85]{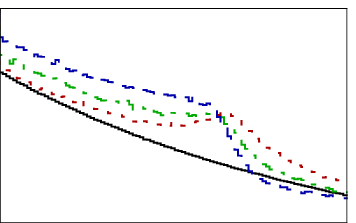} &
\includegraphics[scale=0.85]{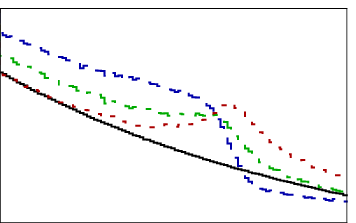} &
\includegraphics[scale=0.85]{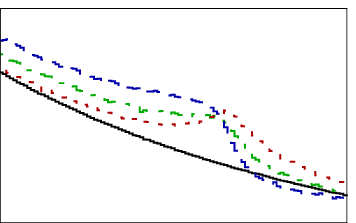} \\
\raisebox{0.35in}[0pt]{
\begin{tabular}{c}
$\phi_\ell=\frac{\pi}{4}$ \\
$g_\ell\propto {1+\gamma_5} $ 
\end{tabular}} &
\includegraphics[scale=0.85]{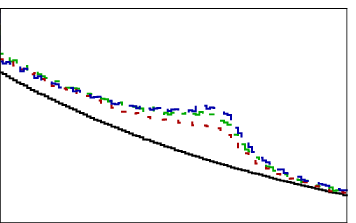} &
\includegraphics[scale=0.85]{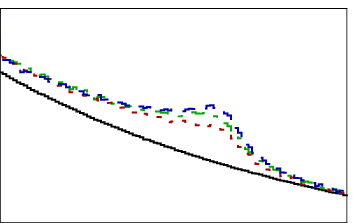} &
\includegraphics[scale=0.85]{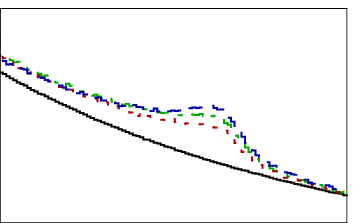} &
\includegraphics[scale=0.85]{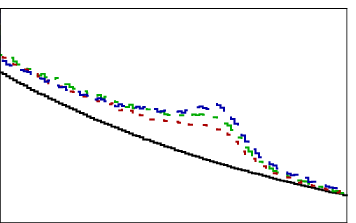} &
\includegraphics[scale=0.85]{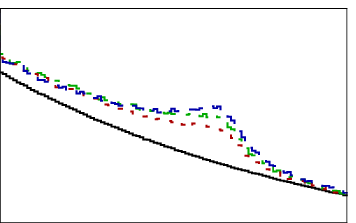} \\
\raisebox{0.35in}[0pt]{
\begin{tabular}{c}
$\phi_\ell=0$ \\
$g_\ell\propto1$ 
\end{tabular}} &
\includegraphics[scale=0.85]{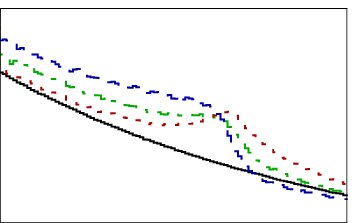} &
\includegraphics[scale=0.85]{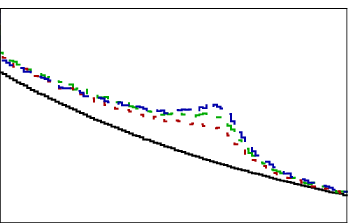} &
\includegraphics[scale=0.85]{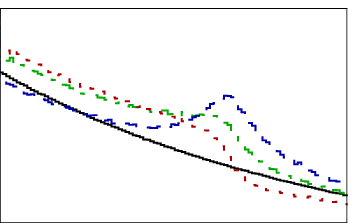} &
\includegraphics[scale=0.85]{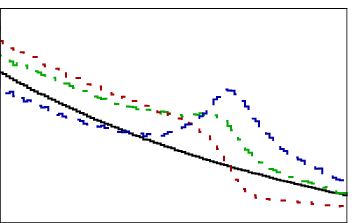} &
\includegraphics[scale=0.85]{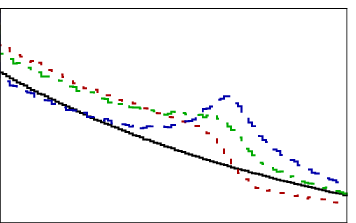} \\
&
$\phi_u=0$ &
$\phi_u=\frac{\pi}{4}$ &
$\phi_u=\frac{\pi}{2}$ &
$\phi_u=\frac{3\pi}{4}$ &
$\phi_u=\pi$ \\
&
$g_u\propto1$ &
$g_u\propto {1+\gamma_5}$ &
$g_u\propto\gamma_5$ &
$g_u\propto {-1+\gamma_5}$ &
$g_u\propto-1$ \\
\end{tabular}
\caption{
(Color online)  
Transverse mass distribution for the process $pp\rightarrow \ell^\pm \nu X$.   The vertical axis is the differential cross section in arbitrary units and the horizontal axis is the transverse mass running from  900~GeV to 1050~GeV. The row and column headers specify the nature of the chiral couplings.  The curve legends, mass, width, energy and pdf set are the same as in Fig. \ref{fig:invmass}.  
\label{fig:transversemass}}
\end{center}
\end{figure*}

The first place we will look for a charged boson in Drell-Yan processes is in the transverse mass.  Since we can not detect the neutrino, we can not fully reconstruct the invariant mass of the lepton system.  The best one can do is to construct the transverse mass which contains the charged lepton momentum and the missing transverse momentum:
\begin{equation}
M_T^2 =
\left(E_{T\ell}+E_{Tmiss}\right)^2-\left({\bf p}_{T\ell}+{\bf p}_{Tmiss}\right)^2~,
\end{equation}
where the transverse energy is defined as $E_T=\sqrt{M^2+p_T^2}$, and $p_{Tmiss}$ is identified as $p_{T\nu}$. 
In practice, we assume that the SM leptons are massless.
Furthermore, in a $2\rightarrow2$ process in the absence of transverse motion,
the missing transverse momentum is equal and opposite to that of the charged lepton. This allows us to simplify the transverse mass to
$M_T=2p_{T\ell}$.
%
This transverse mass distribution develops a Jacobian peak at the
mass of the resonance particle.  We
plot examples of transverse mass distributions for a variety of
parity violation cases in Fig.~\ref{fig:transversemass}. Once again, we see that the shape of the transverse mass distribution depends on the parity properties of the fermionic couplings\footnote{The importance of the transverse mass in determining the properties of the couplings has also been discussed in \cite{Rizzo:2007xs}.}.  Both the vector and the tensor shapes are modified, due to the interference effects near the resonance peak, while the scalar shape remains fixed.  This could be important in determining the mass of a new charged resonance.  Moreover, the discovery of interference in the transverse mass distribution implies that the spin of the new resonant particle is greater than 0 while the sign and size of the interference can give information about the sign and size of the parity violation in the couplings.  We emphasize that the analysis of the transverse mass distribution does not require a knowledge of the quark moving direction, nor the reconstruction of the CM frame.

\subsection{Angular distribution
\label{sec:vprime-angledist}}

\begin{figure}[!tbh]
\begin{center}
\includegraphics{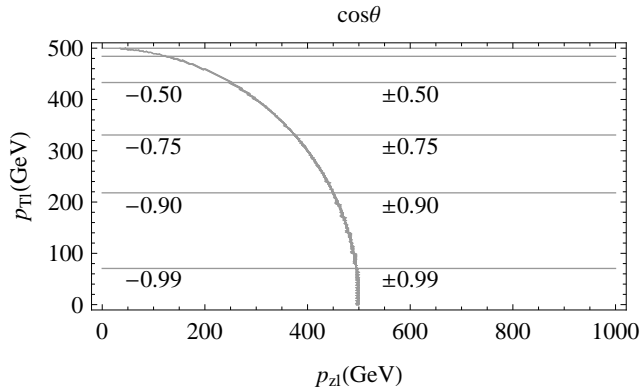}
\end{center}
\caption{\label{cosTheta2sols}Contour plot of $\cos\theta$ as a function of the momentum of the charged lepton
 for the process $pp\rightarrow \ell \nu$.  The horizontal lines are the contours.
The contours are (starting from the bottom) $\pm0.99$, $\pm0.90$, $\pm0.75$, $\pm0.50$, $\pm0.25$ and $0$.  The large solution is negative everywhere while the small solution changes sign at the transition point $E_\ell=M/2$ indicated by the circular curve.}
\end{figure}

%
Unlike the case  mediated by a neutral boson, the angular analysis is known to be difficult for the charged boson mediation 
due to the missing neutrino in the final state, especially for the LHC as a symmetric $pp$ collider\footnote{Other attempts to reconstruct the spin with a missing particle can be found in \cite{Wang:2008sw,Barr:2005dz,Cousins:2005pq,Wang:2006hk,Smillie:2006cd,Barr:2004ze,Smillie:2005ar,Datta:2005zs,Athanasiou:2006ef,Eboli:2011bq,MoortgatPick:2011ix,Cheng:2010yy,Edelhauser:2010gb,Boudjema:2009fz,Gedalia:2009ym,Antipin:2008hj}.}.
%
%
Assuming that
${\bf p}_{T\nu} = -{\bf p}_{T\ell}$,
%
we can solve for $p_{z\nu}$ in terms of the measured charged lepton momentum and the leptonic invariant mass $M_{\ell\nu}$
\begin{equation}
p_{z\nu}
=p_{z\ell}\left(\frac{M^2_{\ell\nu}} {2p_{T\ell}^2}-1\right)\pm
\frac{M_{\ell\nu} E_\ell}{p_{T\ell}}\sqrt{\frac{M^2_{\ell\nu}} {4p_{T\ell}^2}-1}  \  ,
\label{eq: p_znu}
\end{equation}
%
where all quantities are in the lab frame.
If the width of the new resonance is sufficiently small, then the invariant mass is well approximated by the resonant mass
$ M_{\ell\nu} \sim M$.
%
With a clear signal identification, the resonance mass could be measured in the transverse mass distribution, just like the
$M_W$ determination in the SM.
Even so, we still have a two-fold ambiguity in $p_{z\nu}$.  On an event-by-event basis, we do not know which one is correct.
Instead,  for each solution, we can calculate the angle  of the charged lepton in the CM frame with respect to the boost direction
to approximate the quark moving direction, formally defined by
\begin{equation}
\cos\theta = \mbox{sign}(p_{z\ell} + p_{z\nu} )  \ {p_{z\ell}^{CM} \over E_{\ell}^{CM}} ,
\end{equation}
which in turn also suffers from the above ambiguity.
We denote the smaller solution of $|p_{z\nu}|$
by the subscript $S$ and the larger solution by the subscript $L$.
The cosines of these angles
in the CM frame,  can be expressed by the lab quantities as
\begin{eqnarray}
\cos\theta_S &=& -\sqrt{1-\frac{4p_{T\ell}^2}{M^{2}}}\ \mbox{sign}\left(\frac{{M}}{2}-E_\ell\right) , \nonumber\\
\cos\theta_L &=& -\sqrt{1-\frac{4p_{T\ell}^2}{M^{2}}} .
\end{eqnarray}
When $E_{\ell} = M/2$, this corresponds to  the situation where the
lab frame and the CM frame coincide. Both solutions of $\cos\theta$
are identical when $E_\ell \le M/2$ \footnote{However, there is still the possibility that the boost direction does not coincide with the quark direction as in the neutral current case.}.
On the other hand, the two solutions differ by a sign when $E_\ell>M/2$.  In this latter case,
when a wrong solution is chosen, it simply moves the angle from
$\cos\theta\rightarrow-\cos\theta$.  We show a contour plot of
$\cos\theta$ in the $p_{z\ell}-p_{T\ell}$ plane for the two
solutions in Fig.~\ref{cosTheta2sols}.  As is clear in the plot,
$\cos\theta$ does not depend on $p_{z\ell}$ except at the transition
point $E_\ell=M/2=\sqrt{p^{2}_{T\ell} + p^{2}_{z\ell}}$, where it
flips sign.

\begin{figure}[!tb]
\begin{center}
\begin{tabular}{cc}
\multicolumn{2}{c}{Scalar}\\
\includegraphics[scale=0.99]{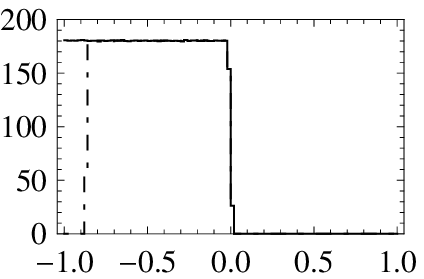} &
\includegraphics[scale=0.99]{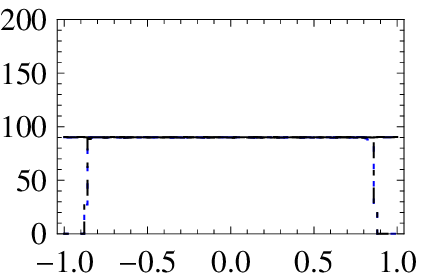}\\
\multicolumn{2}{c}{Vector}\\
\includegraphics[scale=0.99]{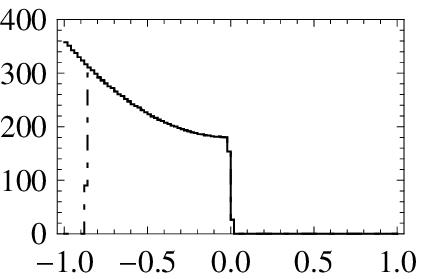} &
\includegraphics[scale=0.99]{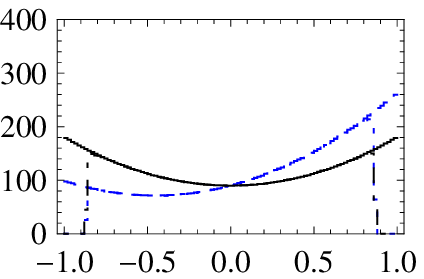}\\
\multicolumn{2}{c}{Tensor}\\
\includegraphics[scale=0.99]{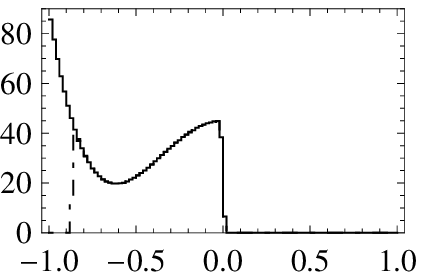} &
\includegraphics[scale=0.99]{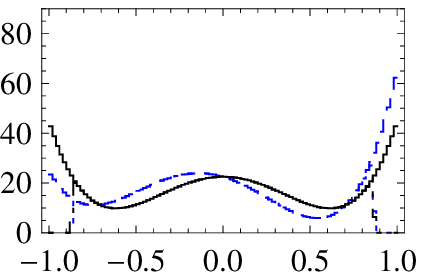}\\
\end{tabular}
\end{center}
\caption{\label{fig:Large sol reconstruction}
(Color online)
Angular distribution of $\ell$  for the process $pp\rightarrow \ell \nu$ 
when only the large solution of the neutrino $p_z$ is taken.
The cut-off feature in the forward-backward regions  is due to the kinematical cuts  as described in the text. }
\end{figure}

These observations lead us to reconstruct adequate angular variables to compensate the loss of information due to the missing
neutrino. By taking the large solution $\cos\theta_{L}$, which is negatively definite, we obtain the distribution $-|\cos\theta|$.
Convoluting with the PDF and imposing the acceptance cuts
\begin{equation}
p_{T\ell} >250\ {\rm GeV}, \quad p_{T miss} >250\ {\rm GeV}, \quad  | \eta_{\ell} | < 2.5,
\label{eq:lncut}
\end{equation}
we reconstruct the angular distributions 
as seen in the left column of Fig.~\ref{fig:Large sol reconstruction}.
We further symmetrize the distribution
by splitting the large solution in half and taking the mirror image on the $\cos\theta>0$ side,
as seen in the right column of Fig.~\ref{fig:Large sol reconstruction} with the solid (black) curves.
With this prescription, we reproduce the correct angular distribution if the true distribution is symmetric ({\it e.g.}, no parity violation),
and we only obtain the average over the positive and negative regions of  $\cos\theta$, as compared with
the dashed (blue) curves in the right column that present left-left chiral couplings for a vector and a tensor state.
%
%
In all cases, the spin information is well preserved and this gives us an unambiguous way to determine the spin of a charged
boson resonance without the need to reconstruct the CM frame.

%

On the other hand, the small solution carries both signs and thus contains information  not only  about the spin, but also the asymmetry.
In the region $E_{\ell}<M/2$, this solution is the same as the large solution and thus the  angle is typically correctly reconstructed.
In the region $E_{\ell}>M/2$, however, either this solution or the one with an opposite sign could be right and we cannot determine it on
an event-by-event basis.
In each plot in the left column of Fig.~\ref{fig:Small sol
reconstruction}, we plot the angular distribution using the small
solution. The distribution is different for each spin and asymmetry
and it generally covers the entire range of $\cos\theta$.
%
%
As expected, the small solution is essentially correct about $\cos\theta \lesssim -0.6$,
but there is a clear deficit when $\cos\theta$ approaches $0^{-}$.
The missing events are incorrectly assigned to the bins in $-\cos\theta>0$, which is also seen as an excess near $\cos\theta \to 0^{+}$.

Fortunately, we are able to simulate the expectation and thus restore the distribution on a statistical basis.
In each plot on the right column of Fig.~\ref{fig:Small sol
reconstruction}, the dashed (blue) curve is the expected angular
distribution if we know the full momentum information of the neutrino.
%
%
For $\cos\theta > 0$,
we take the difference between the solid (black) curve on the left column and the dashed (blue) curve  on the right column,
to obtain the excess. We would like to move those events back to the bins in $-\cos\theta < 0$ to restore the original distribution.
It is very important to realize that the fractional excess (the above excess divided by the solid (black) curve on the left column)
turns out to be numerically the same for all of the scalar, vector and tensor resonances. 
Denoted by $f$ as a function of $\cos\theta$, we obtain this simulated fractional excess as a universal function for all spins,
shown in Fig.~\ref{cosThetaFraction}. 
This observation leads to a powerful procedure for the restoration of the angular distribution: 
Given a data set, presumably like the solid (black) curve on the left column, we apply the fractional excess function to the
$\cos\theta > 0$ region bin by bin, then subtract this result out from the data, and finally move the result to the region of 
$\cos\theta < 0$ for the correction. 

Plotted on the right column in solid (black) are the corrected angular distributions.  In the case of symmetric distributions, the reconstruction works perfectly and the solid (black) curve completely coincides with the expectation of the dashed (blue) line.  In the case of asymmetric distributions, the reconstruction is perfect on the edges but slightly off in the middle where the dashed (blue) curve can be seen.
This is due to the more likely mismatch between the directions for the true quark momentum and the boost.
Nevertheless, it is sufficient to present features of the asymmetry.
%
%
We reiterate that our correction for each distribution is independent of the resonance spin, 
although it depends on the parton distribution function as well as the mass of the new resonant particle.

%
%
%
%
%

\begin{figure}[!tb]
\begin{center}
\begin{tabular}{cc}
\multicolumn{2}{c}{Scalar}\\
\includegraphics[scale=0.99]{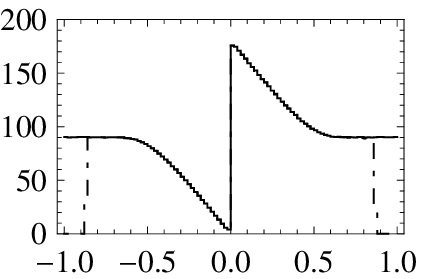} &
\includegraphics[scale=0.99]{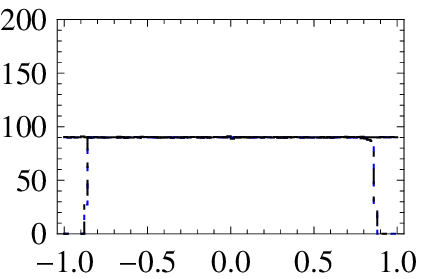}\\
\multicolumn{2}{c}{Symmetric Vector}\\
\includegraphics[scale=0.99]{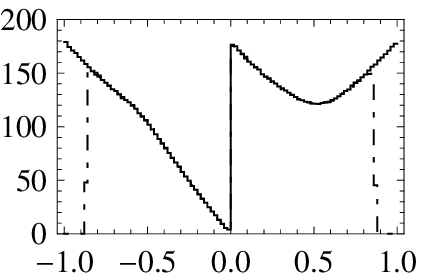} &
\includegraphics[scale=0.99]{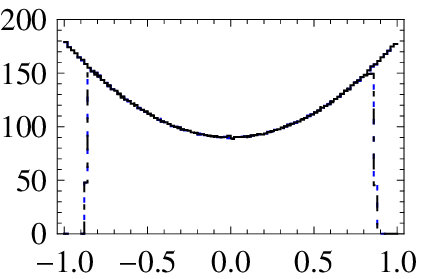}\\
\multicolumn{2}{c}{Left-Left Vector}\\
\includegraphics[scale=0.99]{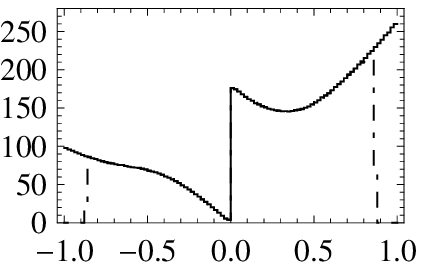} &
\includegraphics[scale=0.99]{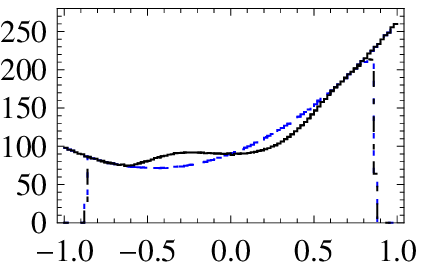}\\
\multicolumn{2}{c}{Symmetric Tensor}\\
\includegraphics[scale=0.99]{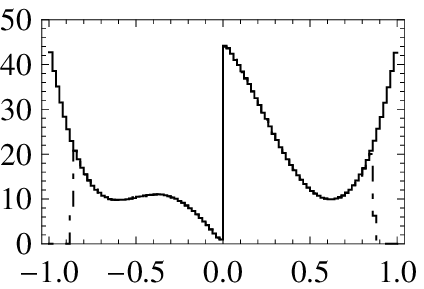} &
\includegraphics[scale=0.99]{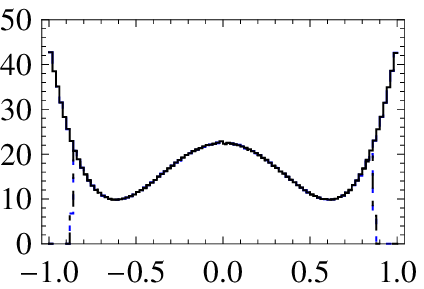}\\
\multicolumn{2}{c}{Left-Left Tensor}\\
\includegraphics[scale=0.99]{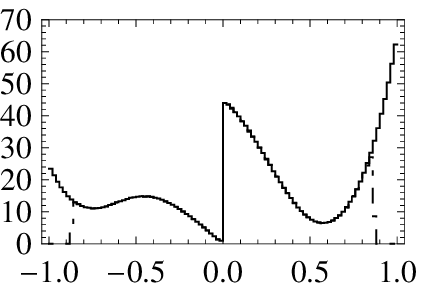} &
\includegraphics[scale=0.99]{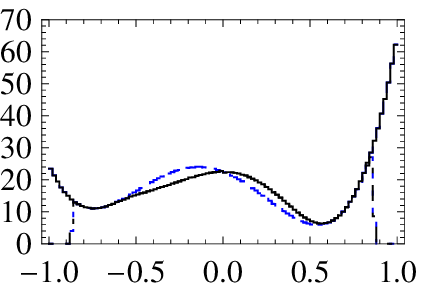}\\
\end{tabular}
\end{center}
\caption{\label{fig:Small sol reconstruction}
(Color online)
Angular distribution of $\ell$ for the process $pp\rightarrow \ell \nu$ 
when only the small solution of the neutrino $p_z$ is taken.  Each row lists a case of distinct spin and parity violation.  Each plot on the left gives the angular distribution directly obtained from the small solution.  On the right, the dashed (blue) curve is the angular distribution with full momentum information of the neutrino.  The solid (black) curve is the angular distribution obtained from the corresponding distribution on the left by following the procedure explained in the main text.
The cut-off feature in the forward-backward regions  is due to the kinematical cuts  as described in the text.
}
\end{figure}


\begin{figure}
\begin{center}
\includegraphics{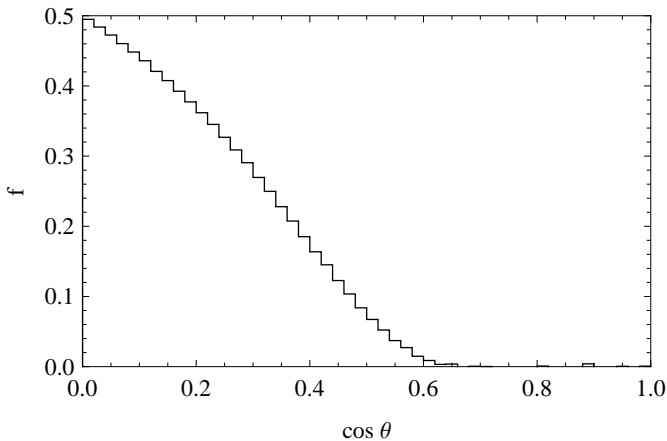}
\end{center}
\caption{\label{cosThetaFraction}
Fraction of events that should be removed from the $\cos\theta > 0$ 
side of the small-solution angular distribution and added to the $\cos\theta < 0$ side of the distribution, 
universal for a charged resonance of any spin.}
\end{figure}

%
%
%

\subsection{Forward-backward asymmetry
\label{sec:vprime-fba}} The angular distributions for the large
solution average out the events and thus lead to no asymmetry. The
corrected angular distributions for the small solution as in the
right column of Fig.~\ref{fig:Small sol reconstruction} preserve the
asymmetry property to a large extent. We could use these angular
distributions in the same way as in the neutral current case, or we
could use the small solution directly only in the region
$|\cos\theta| \gtrsim 0.5$ without performing the $\cos\theta
\leftrightarrow -\cos\theta$ correction. Thus similar analyses to
Fig.~\ref{fig:FBA1} in the the neutral currents can be performed.
This is a significant progress for the charged boson signal at the
LHC as a symmetric collider.


\section{Searches for Drell-Yan Type Signals at the LHC
\label{sec:Search}}

\subsection{\label{sec:neutral CMS}Neutral current channel}

The CMS and ATLAS Collaborations have studied dilepton events at $7$~TeV in search of a heavy neutral 
boson \cite{Chatrchyan:2011wq,Aad:2011xp}.
They present their results in terms of the production cross section times branching fraction as a function of mass, which takes into account the couplings, mass and width of the new particle for general spin.  In this note, we extend this by projecting the expected $95\%$ confidence level bounds to 10 fb$^{-1}$ at $7$~TeV and 10 fb$^{-1}$ and 300 fb$^{-1}$ at $14$~TeV.  We also place 
bounds on the product of couplings as a function of mass for each spin while using a standard width.  

\begin{figure}[!tb]
\begin{center}
\begin{tabular}{cc}
\includegraphics[scale=1]{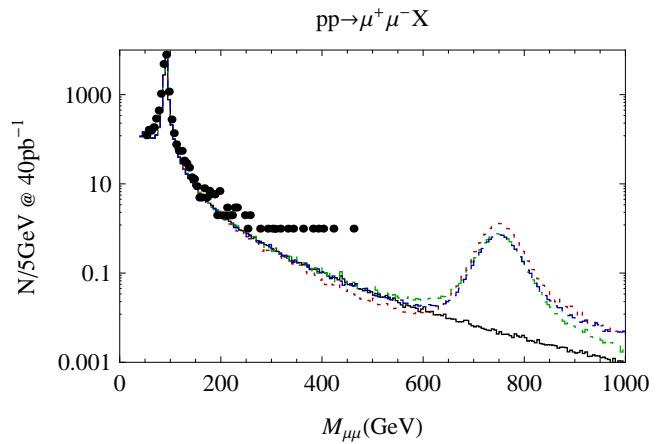} 
\end{tabular}
\caption{\label{fig:Meedist}
(Color online)  
Invariant mass distribution of the charged leptons  for the process $pp\rightarrow \mu^+\mu^-X$ 
at the LHC with $7$~TeV.  The solid (black) curve is the SM expectation while the dot-dashed (green), dotted (red) and dashed (blue) curves are respectively for spin-0, -1, and -2 resonances with a mass of 750~GeV and a width of 22.5~GeV.  The latest CMS data \cite{Chatrchyan:2011wq} are
superimposed in the figure as the (black) dots.  
}
\end{center}
\end{figure}

We simulate the process $pp\rightarrow \ell^+ \ell^- X$ in the SM at tree level using the parton-level Monte Carlo package CalcHEP \cite{Pukhov:2004ca,Pukhov:1999gg} and normalize our parton-level calculations to the data by fitting to the two highest bins (from 85~GeV to 95~GeV) from Fig.~2 of Ref.~\cite{Chatrchyan:2011wq}.  
%
Including the numerical normalization factors 0.56 for the $e^+e^-$ channel and 1.01 for the $\mu^+\mu^-$ channel,
our results and the CMS data are 
given in Fig.~\ref{fig:Meedist} by the solid (black) curve and the (black) dots, respectively.
Signals, including interference with the SM background, corresponding to new resonances of different spins, masses, and fermionic couplings were then done and superimposed on the plot.  For simplicity and illustration purposes, we assumed that all the couplings were taken to be real in the numerical studies and that only the first two generations of fermions were taken into account.  Results for more general cases are expected to be qualitatively the same.

\begin{figure}[!tb]
\begin{center}
\includegraphics[scale=1.2]{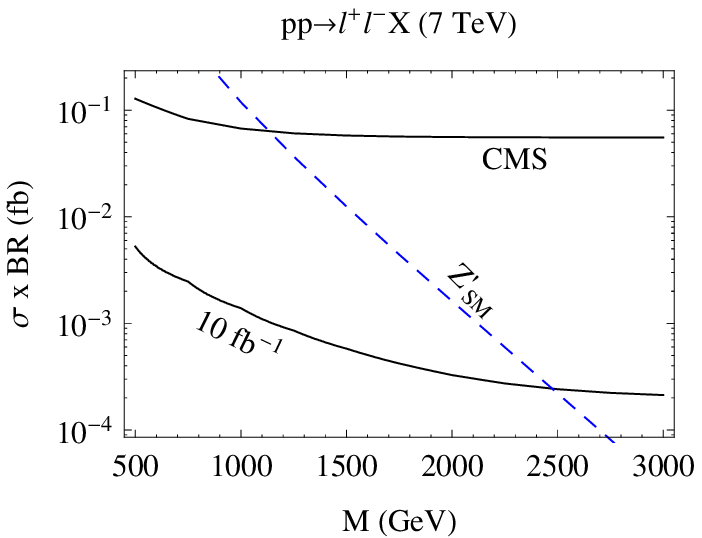}\\
\vspace{0.1in}
\includegraphics[scale=1.2]{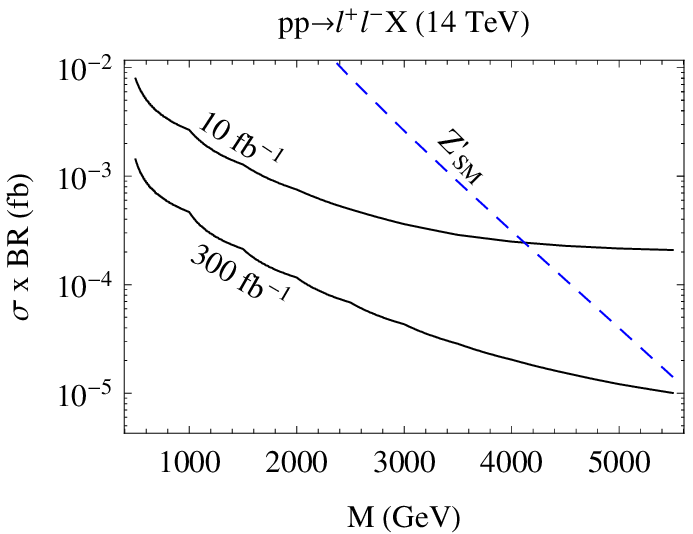}
\caption{\label{fig:neutrallowerbound}
(Color online)  
95\% C.~L. bound on the production cross section times branching fraction versus the new resonant boson mass.  The curve labeled CMS is with respect to the data measured by the CMS Collaboration \cite{Chatrchyan:2011wq}, while the other solid (black) curves are projections for the specified integrated luminosity. 
The dashed (blue) curve is for a $Z'$ with the same couplings as the SM $Z$.
}
\end{center}
\end{figure}

In numerical estimates of the signal events, the width of the new resonance was taken as 3\% of its mass, in line with the assumption made in Ref.~\cite{Chatrchyan:2011wq}.  We considered the events in the invariant mass window of $\pm 20\%$ of its mass
\begin{equation}
\frac{4}{5}M \le M_{\ell\ell} \le \frac{6}{5}M ~,
\end{equation}
where $M$ is the mass of the new particle.  The number of events for the $e^+e^-$ and $\mu^+\mu^-$ channels were added ($N=N_e+N_\mu$) to give the total number of predicted events for the SM. 
This number of events was plugged into the one bin log likelihood \cite{Nakamura:2010zzi}
\begin{equation}
LL = 2 \left[
N \ln \left(\frac{N}{\nu}\right) + \nu - N
\right] ~,
\end{equation}
where $\nu$ is the number of events (for both the $e^+e^-$ and $\mu^+\mu^-$ channels) expected in the SM plus the new boson.
A value of $LL=4$ was taken as the 95\% confidence level and $\nu$ was solved for.
Taking into account the SM expectation and given an available integrated luminosity, this is converted into a signal cross section times the branching fraction for a given mass, and the results 
are plotted in Fig.~\ref{fig:neutrallowerbound} for $7$~TeV and $14$~TeV.

We wish to emphasize that our results here can be broadly applied to any resonant signal as outlined in the earlier sections.
The production cross section is governed by the resonant coupling to the initial state partons, and the decay branching fraction
is proportional to the coupling to the final state leptons. Thus with the determination of the resonant mass and width by the
kinematical peak and the line shape, we expect to gain the information for the fundamental couplings $g_{q}g_{\ell}$ for a 
scalar or a vector resonance, and $g_{q}g_{\ell}/\Lambda^{2}$ for a tensor. 

To illustrate this point, we take the commonly studied ``sequential $Z'$ model'' as an example for a vector resonance ($Z'_{SM}$), 
which has the same coupling as the SM $Z$ boson. 
In Fig.~\ref{fig:neutrallowerbound}, we also 
include the cross section times branching fraction (the blue dashed curve) for this model as labeled by $Z'_{SM}$. 
Where this curve crosses the solid (black) curve gives the bound (or projected bound) for this model.  We find a lower bound on $Z'_{SM}$ of $1135$~GeV consistent with the CMS results.  We also find that CMS could bound the mass at $\sim2.5$~TeV with 10 fb$^{-1}$ at $7$~TeV, $\sim4.1$~TeV with 10 fb$^{-1}$ at $14$~TeV and $\sim5.5$~TeV with 300 fb$^{-1}$ at $14$~TeV.   This procedure can easily be applied to other specific models.  The couplings and widths simply need to be set and the mass scanned over to determine the cross section times branching fraction.

%
%
%
%
%
%
%

\subsection{\label{CMS analysis}Charged current channel}

\begin{figure}[!tbh]
\begin{center}
\begin{tabular}{cc}
\includegraphics[scale=1]{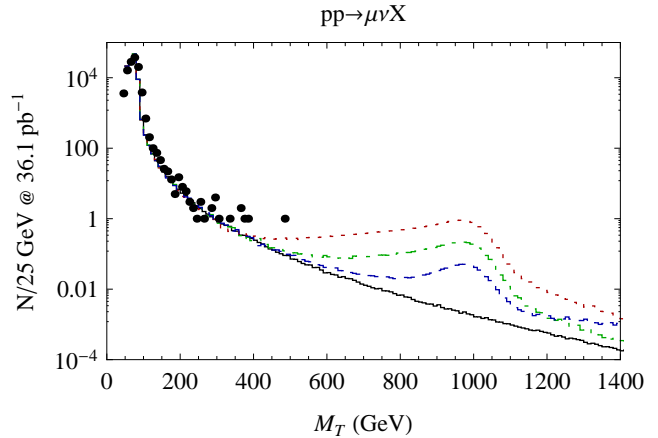} 
\end{tabular}
\caption{\label{fig:MTdist}
(Color online) 
Transverse mass distribution of the charged lepton at the LHC with $7$~TeV.  The solid (black) curve is the SM expectation while the dot-dashed (green), dotted (red) and dashed (blue) curves are respectively for spin-0, -1, and -2 resonances with a mass of 1~TeV and a width of 20~GeV.
The latest CMS data \cite{Chatrchyan:2011dx} are superimposed as the (black) dots.
}
\end{center}
\end{figure}

Following the same procedure as in the previous section, we calculate the 95\% confidence level bounds and projected bounds on a new charged resonant boson.  We base our results on the
the CMS data published in \cite{Chatrchyan:2011dx,Khachatryan:2010fa}. 
We simulate $pp\rightarrow \ell\nu \  (\ell=e,\mu)$ in the SM at tree level, again 
using the parton-level Monte Carlo package CalcHEP,
and normalize our parton-level
calculations to the data by multiplying our simulation with a
numerical factor which brings the two highest bins (from 50~GeV to
100~GeV) into agreement.  We found the numerical factor to be $0.76$
for the electron and $0.91$ for the muon.  A plot of our SM
calculation and the CMS data can be seen in Fig.~\ref{fig:MTdist}
where the solid (black) line is the parton-level prediction of the
SM and the (black) dots are the CMS data for the case of the
muon.

The significance was calculated by taking the number of events in the window
\begin{equation}
\frac{2}{5}M<M_T<\frac{6}{5}M~.
\end{equation}
The one bin log likelihood was then calculated for the sum of the electron case and the muon case.
%
%

\begin{figure}[!tb]
\begin{center}
\includegraphics[scale=1.2]{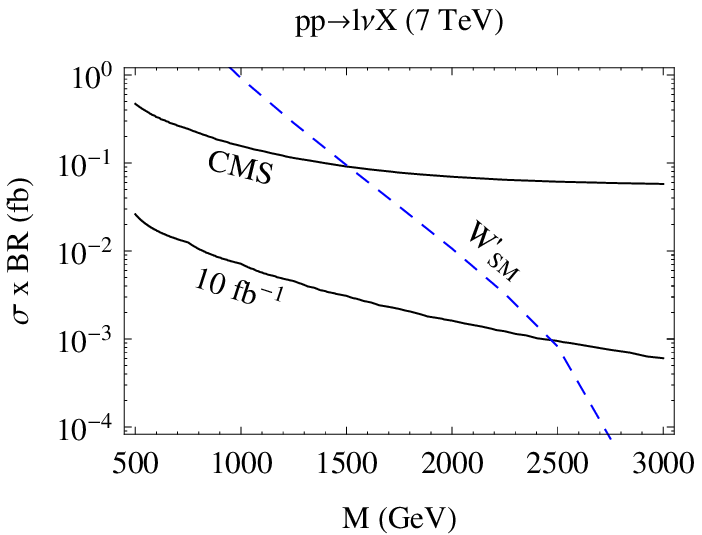}\\
\vspace{0.1in}
\includegraphics[scale=1.2]{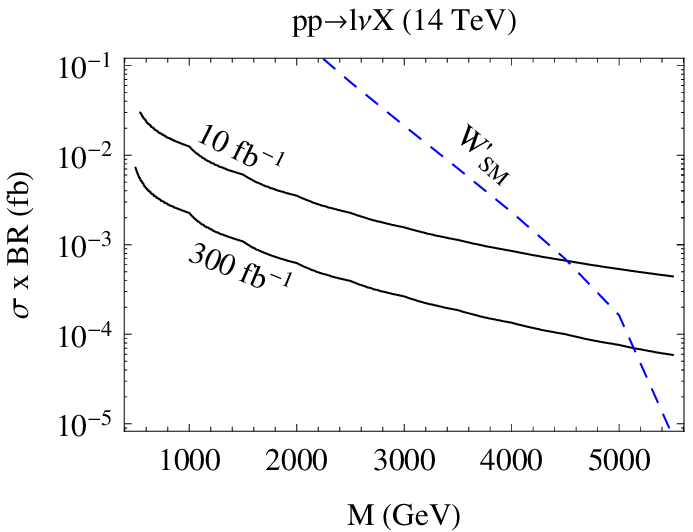}
\caption{\label{fig:lowerbound}
(Color online) 
95\% C.~L. bound on the production cross section times branching fraction versus the new resonant boson mass.  The curve labeled CMS is with respect to the data measured by the CMS Collaboration \cite{Khachatryan:2010fa,Chatrchyan:2011dx}, while the other solid (black) curves are projections for the specified integrated luminosity.  The dashed (blue) curve is for a $W'$ with the same couplings as the SM $W$.}
\end{center}
\end{figure}

We again plotted the $LL=4$ line as the 95\% confidence level in
Fig.~\ref{fig:lowerbound}.  Once again, to illustrate the approach to generalize to other models, 
we plot the cross section times branching fraction of a new charged vector boson with the same couplings as that in the SM, 
the ``sequential  $W'$ model'', denoted by $W'_{SM}$.  We find that $W'_{SM}$ is bound to be heavier than $1500$~GeV at $95\%$ confidence level which is close to the CMS result \cite{Khachatryan:2010fa}.  We also find that CMS could achieve a bound of $\sim2.5$~TeV with 10 fb$^{-1}$ at $7$~TeV, a bound of $\sim4.5$~TeV with 10 fb$^{-1}$ at $14$~TeV and a bound of $\sim5.2$~TeV with 300 fb$^{-1}$ at $14$~TeV.  We again note that this procedure can be followed with any models.  After the couplings and widths are set appropriately, the cross section times branching fraction can be calculated and plotted as a function of mass.
%

%
%
\section{Summary
\label{sec:summary}}
We have considered the most general new resonant $s$-channel contributions to Drell-Yan production of leptons at the LHC including spin-0, 1 and 2 bosons.  We formulated the most general leading-order interactions between these new particles and the SM fields involved in the DY channel that satisfy Lorentz and EM invariances including both parity conserving and violating terms.  Using these interactions, we have calculated the helicity amplitudes and expressed them in terms of the Wigner $d^j_{m,m'}$ functions, explicitly showing the angular dependence of these collisions in the CM frame.

For the neutral current process $pp\rightarrow\ell^+\ell^-X$, we find that
\begin{itemize}
\item 
the lepton pair invariant mass distribution may provide information for the chiral interactions.
A new spin-1 field interferes with the SM process, thus modifying its shape from the Breit-Wigner resonance and the position of its peak,
while both spin-0 and spin-2 fields do not have significant interference with the SM and appear as simple Breit-Wigner resonances.  

\item
the transverse momentum of the charged lepton also provides information for the chiral interactions.  Both the spin-1 and spin-2 fields
present interference with the SM background.  Along with the invariant mass distribution, this gives another way to distinguish between the spins. 

\item 
defining an angle of $\ell^-$ with respect to the boost direction (likely to be the initial quark direction) in the CM frame of the system,
one is able to construct the Wigner $d^j_{m,m'}$ functions as well as their asymmetry due to parity violation.  We note that an asymmetry only occurs for a spin-1 or 2 boson when both the quark coupling and the lepton coupling are not either purely vectorial (tensorial) or axial vector (axial tensor) and is maximized when the vectorial (tensorial) and axial (axial tensor) couplings are equal in magnitude.

\item 
although it is well-known that a chiral spin-1 boson generates an asymmetry when the forward and backward directions are defined in the full angular range from $0<\cos\theta<1$ and $-1<\cos\theta<0$, respectively, the tensor asymmetry would be zero if taking this range.  Instead, we find that the asymmetry is maximized for the tensor when considering the 
asymmetry in the range
$1/\sqrt{2}<\cos\theta<1$ and $-1<\cos\theta<-1/\sqrt{2}$, respectively. Thus the asymmetry gives a new way to distinguish between the spins. 

\item 
scanning over mass and cross section times branching fraction for the new bosons, we obtained the $95\%$ confidence level bounds based on the current CMS results as well as projected bounds for future integrated luminosities and machine energies. To illustrate how to apply our formulation and calculation to other DY resonances,  we plotted the cross section times branching fraction as a function of mass for a new vector boson with the same couplings as the SM $Z$ boson (denoted by $Z'_{SM}$).

\end{itemize}

For the charged current process $pp\rightarrow \ell^\pm\nu X$, we find that
\begin{itemize}
\item 
the $\ell\nu$ transverse mass distribution may provide information for the chiral interactions.  Spin-1 and spin-2 fields interfere with the SM process, thus modifying its shape from the conventional noninterfering transverse mass distribution as well as the position of the Jacobian peak while a spin-0 field does not have significant interference with the SM and appears with the conventional shape.  The shifted position of the peak can be important in determining the mass of the new resonance.

\item 
although determining the angle of $\ell^\pm$ with respect to the boost direction is more challenging due to the two-fold ambiguity in the $z$-component of the neutrino momentum, we find a novel statistical method for reconstructing the angular distribution.  This method involves creating two distributions, one for the small neutrino $z$-component momentum and another for the large neutrino $z$-component momentum.  We find that the large solution faithfully preserves the symmetrized Wigner $d^j_{m,m'}$ function and therefore fully determines the spin of the new resonance but not the asymmetry.  The small solution distribution, on the other hand, does contain information about the asymmetry and we show how that information can be extracted.  

\item 
scanning over mass and cross section times branching fraction for the new bosons, we obtained the 95\% confidence level bounds based on the current CMS results as well as projected bounds for future integrated luminosities and machine energies.  To illustrate how to apply our formulation and calculation to other DY resonances, we plotted the cross section times branching fraction as a function  of mass for a new vector boson with the same couplings as the SM $W$ boson (denoted $W'_{SM}$).

\end{itemize}

We would like to reiterate that our formulation makes the future phenomenological and experimental searches straightforward.
Our proposal for the large and small solutions for the charged current channel overcomes the difficulty of the two-fold ambiguity 
due to the missing neutrinos, that should be adopted for studies of new charged resonances.

\section*{Acknowledgments}

We would like to thank Camilo Garcia for discussions.
C.W.C was supported by NSC Grant No. 97-2112-M-008-002-MY3.    N.D.C. is supported by the US National Science Foundation, grant NSF-PHY-0705682, the LHC Theory Initiative.  G.J.D. is supported by the National Natural Science Foundation of China under
Grant No.10905053.  T.H. is supported in part by the U.S. Department of Energy under grant No.
DE-FG02-95ER40896.  

\appendix
%
%
\section{Feynman Rules
\label{sec:feynman rules}}

\begin{figure*}[hptb]
\begin{tabular}{ll}
\includegraphics[scale=.5]{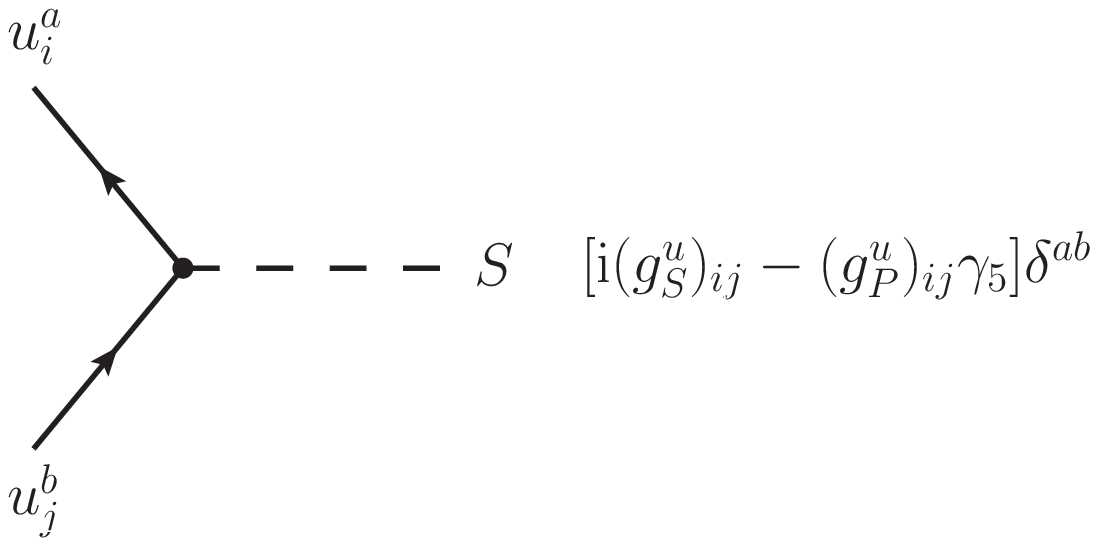}&\hspace{0.5cm}
\includegraphics[scale=.5]{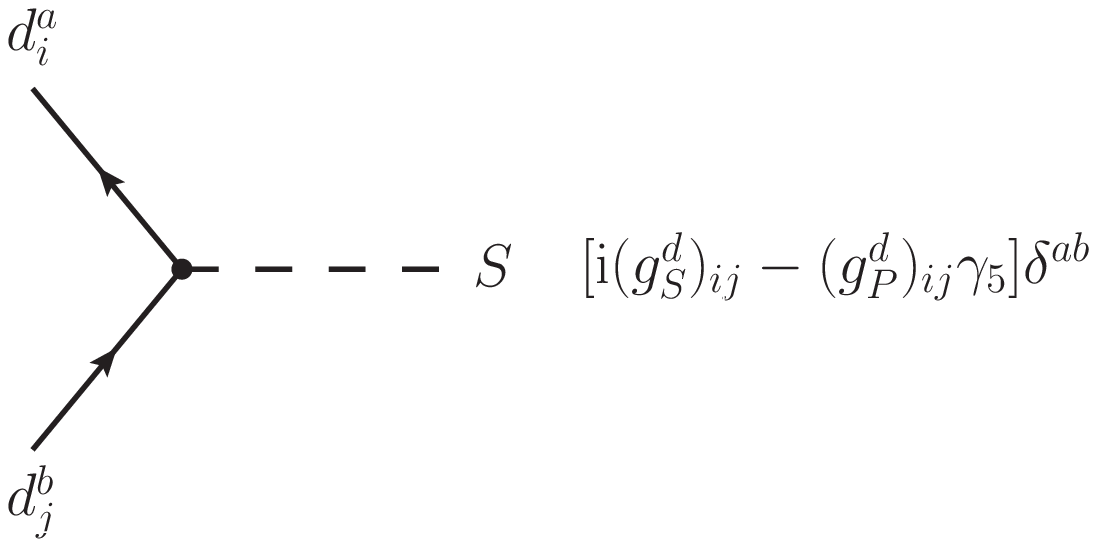}\\
\includegraphics[scale=.5]{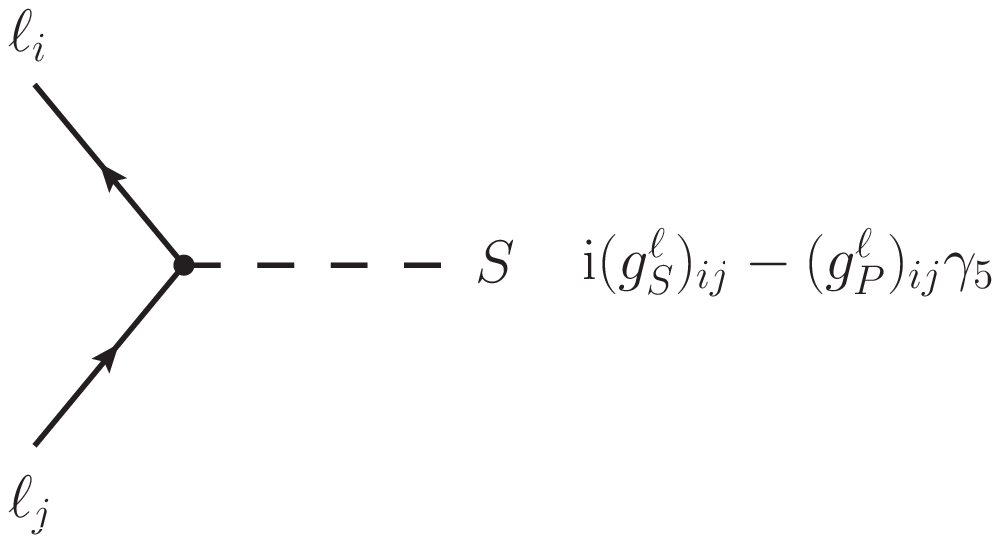}&\hspace{0.5cm}
\includegraphics[scale=.5]{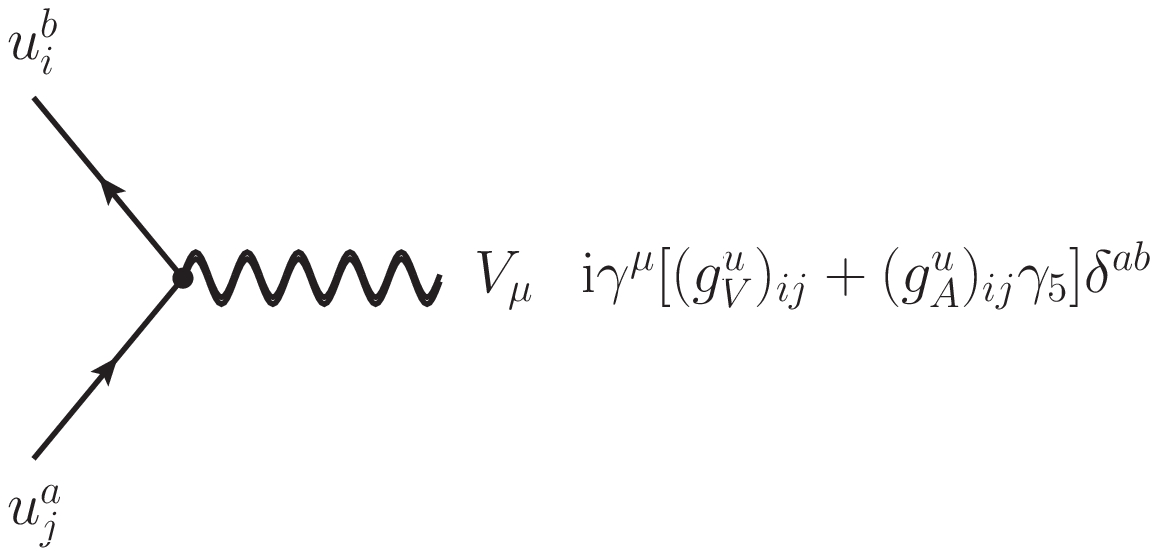}\\
\includegraphics[scale=.5]{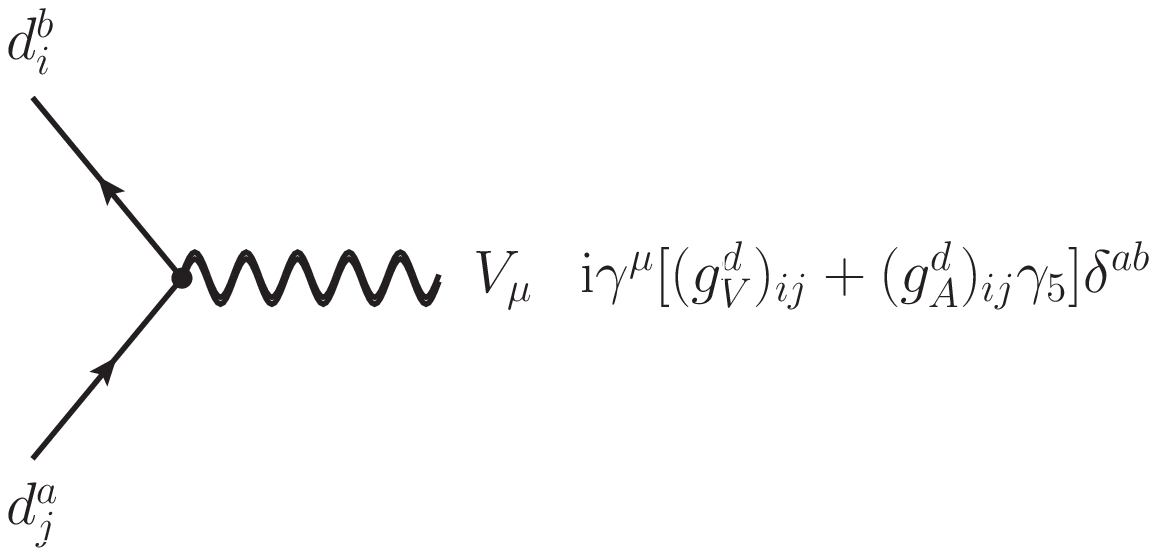}&\hspace{0.5cm}
\includegraphics[scale=.5]{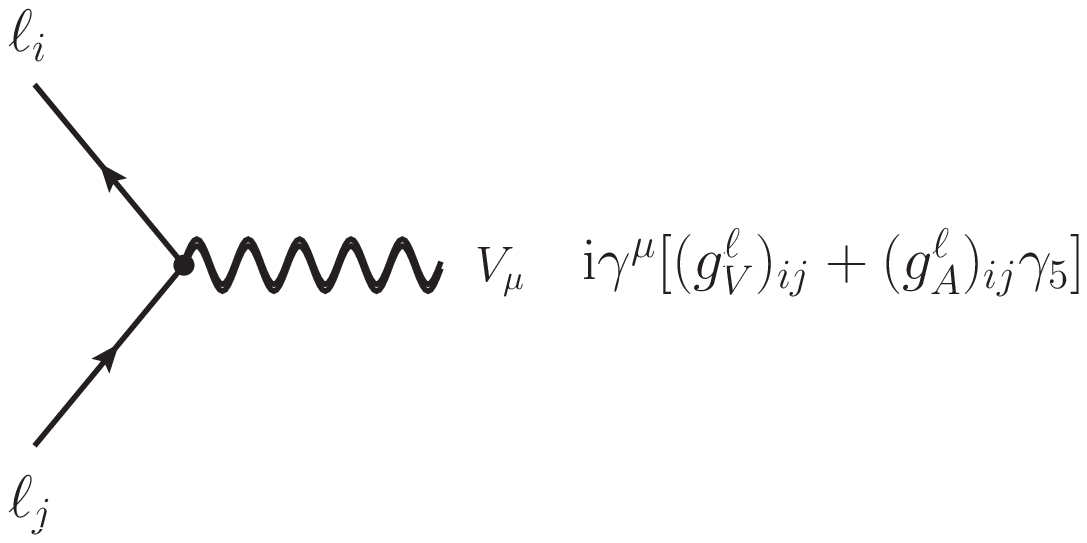}\\
\includegraphics[scale=.5]{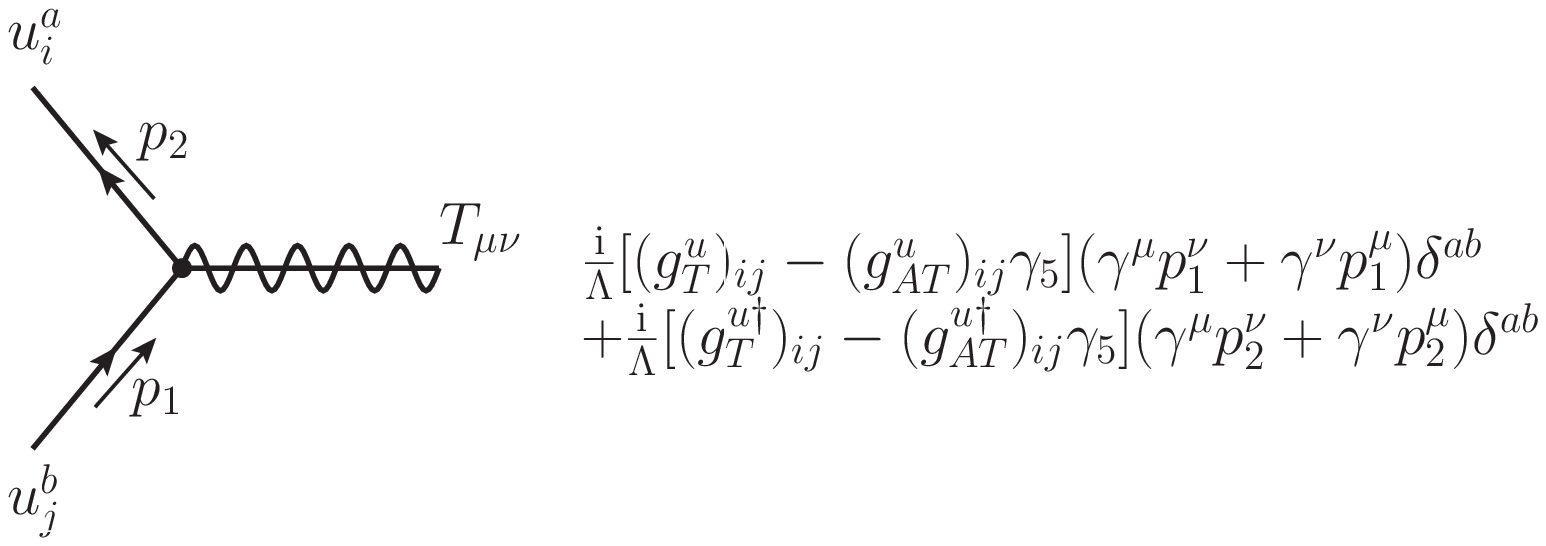}&\hspace{0.5cm}
\includegraphics[scale=.5]{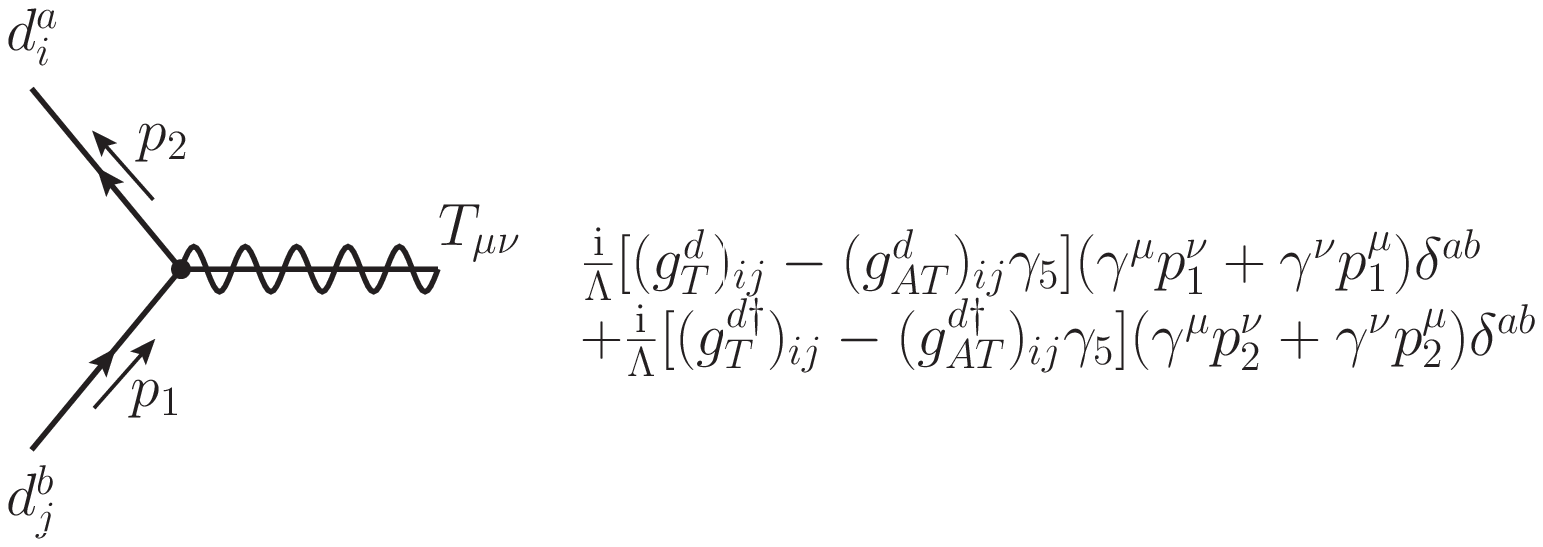}\\
\includegraphics[scale=.5]{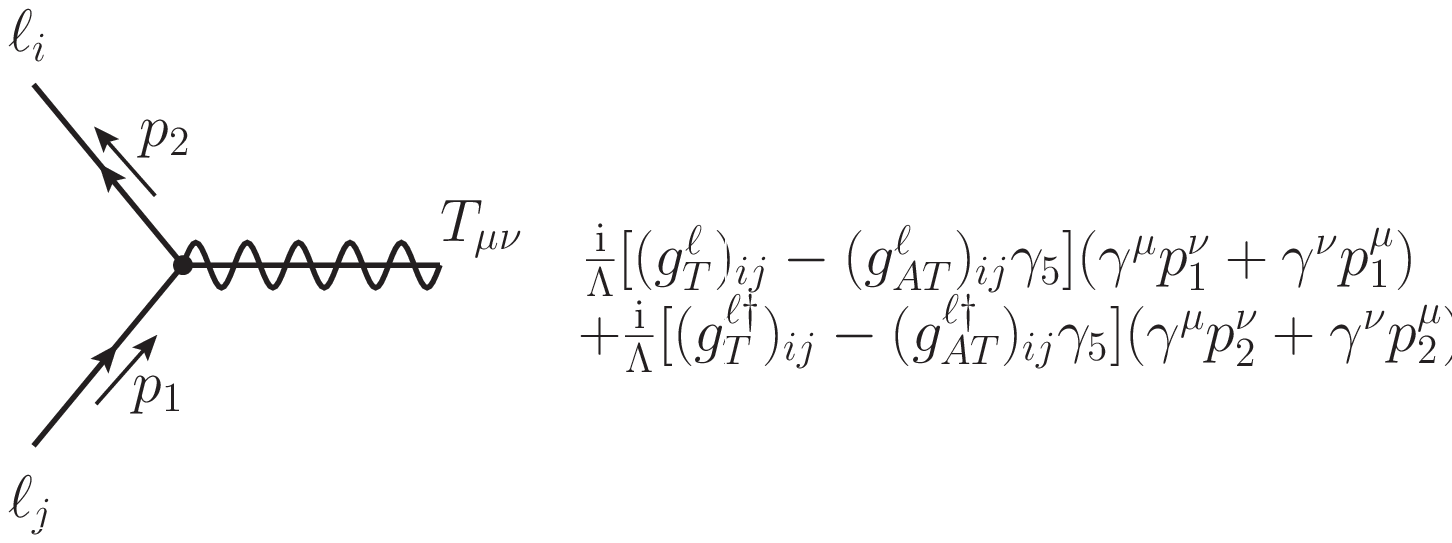}&\hspace{0.5cm}\includegraphics[scale=.5]{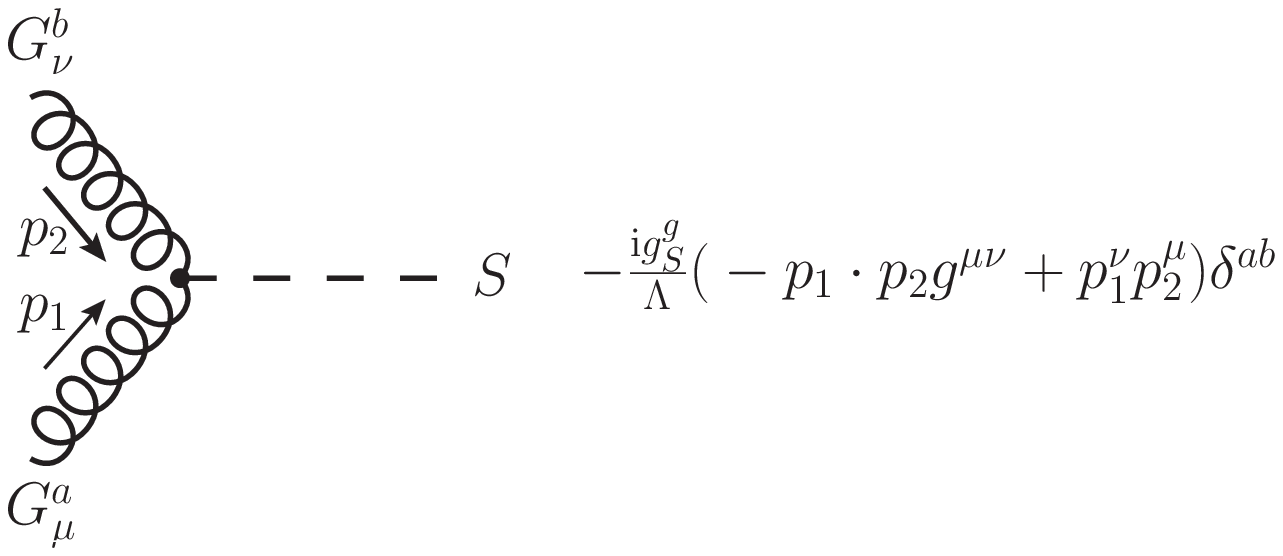}\\
\includegraphics[scale=.5]{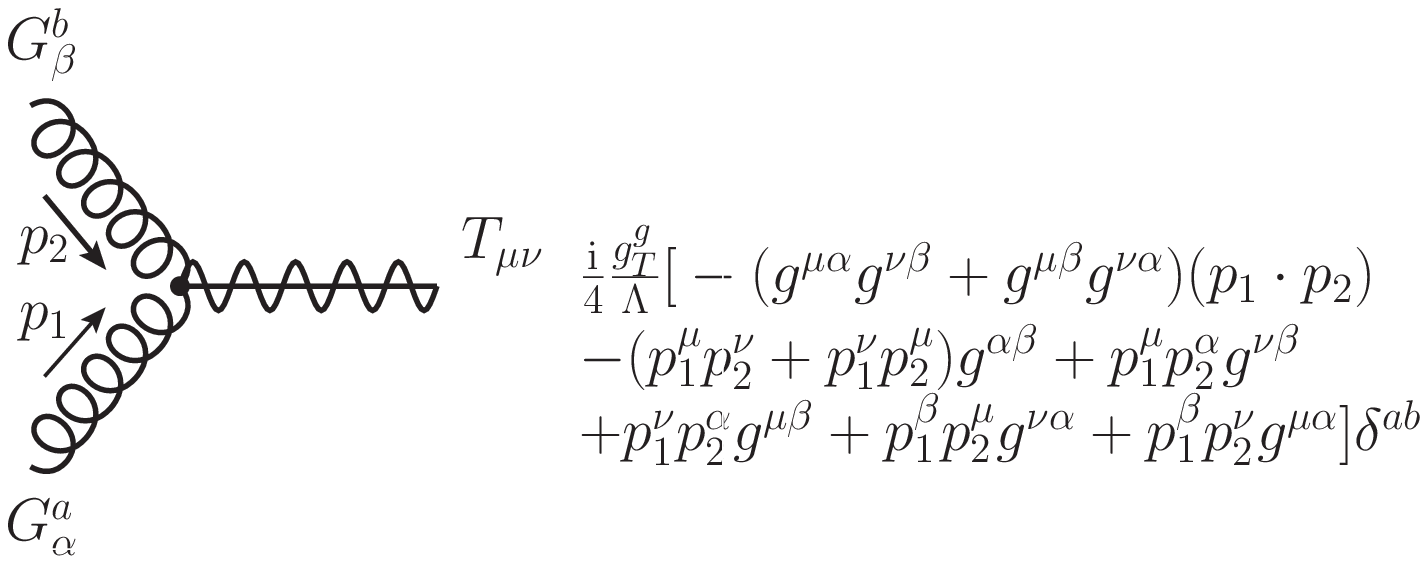}&\hspace{0.5cm}
\end{tabular}
\caption{\label{fig:V} Interaction vertices between the new bosons and SM fields.}
\end{figure*}

\begin{figure*}[hptb]
\begin{center}
\begin{tabular}{ll}
\includegraphics[scale=.5]{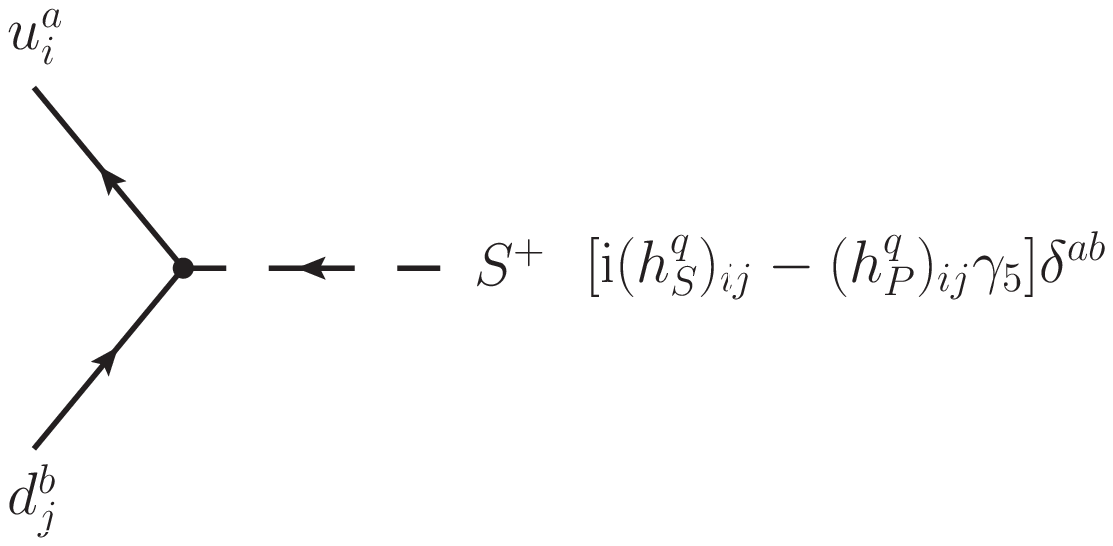}&\hspace{0.5cm}
\includegraphics[scale=.5]{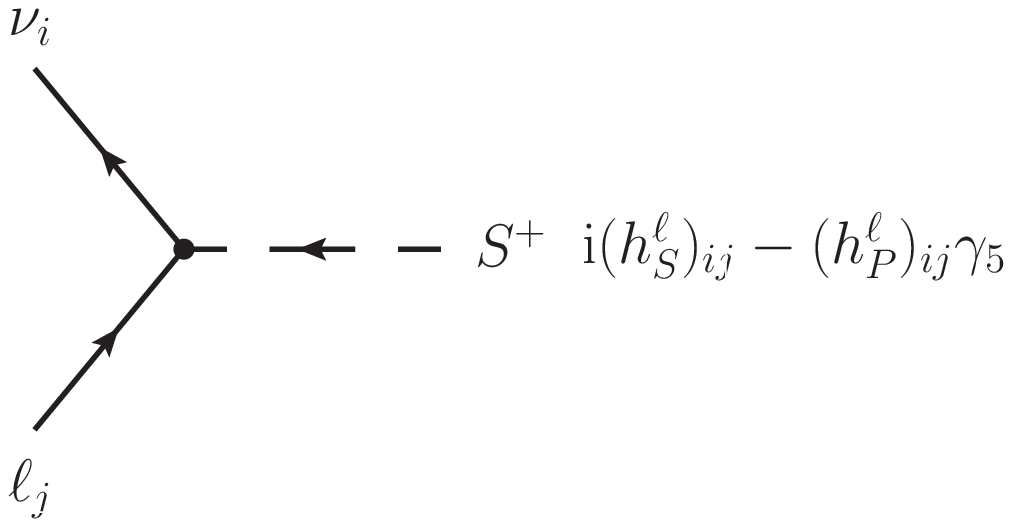}\\
\includegraphics[scale=.5]{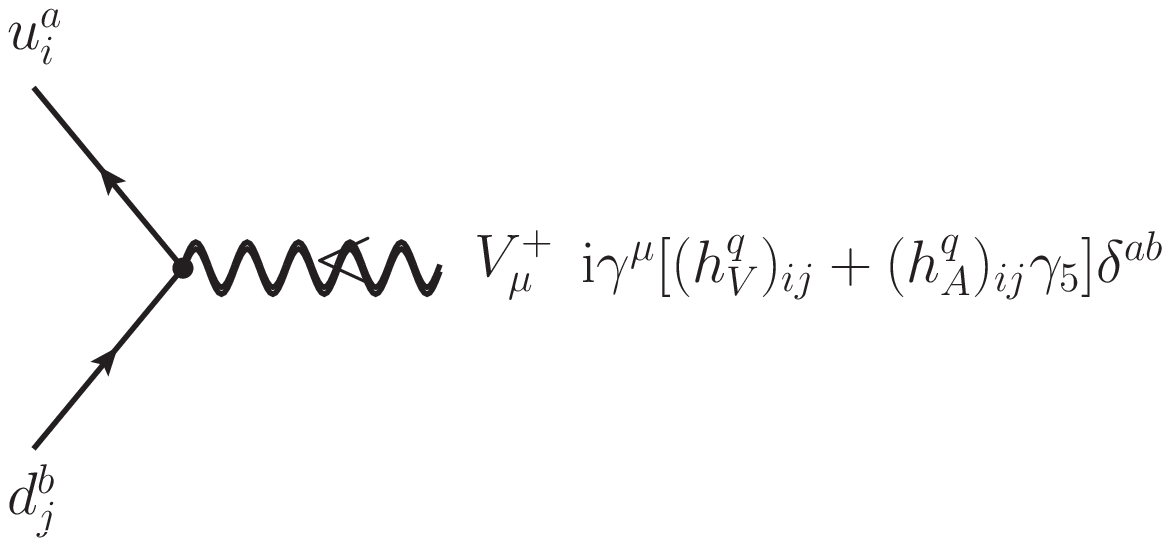}&\hspace{0.5cm}
\includegraphics[scale=.5]{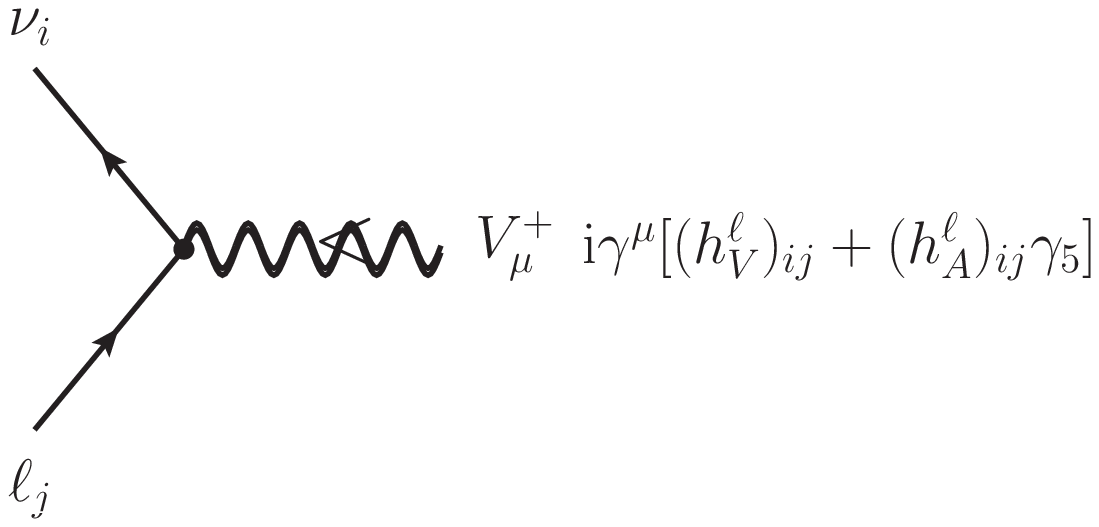}\\\
\includegraphics[scale=.5]{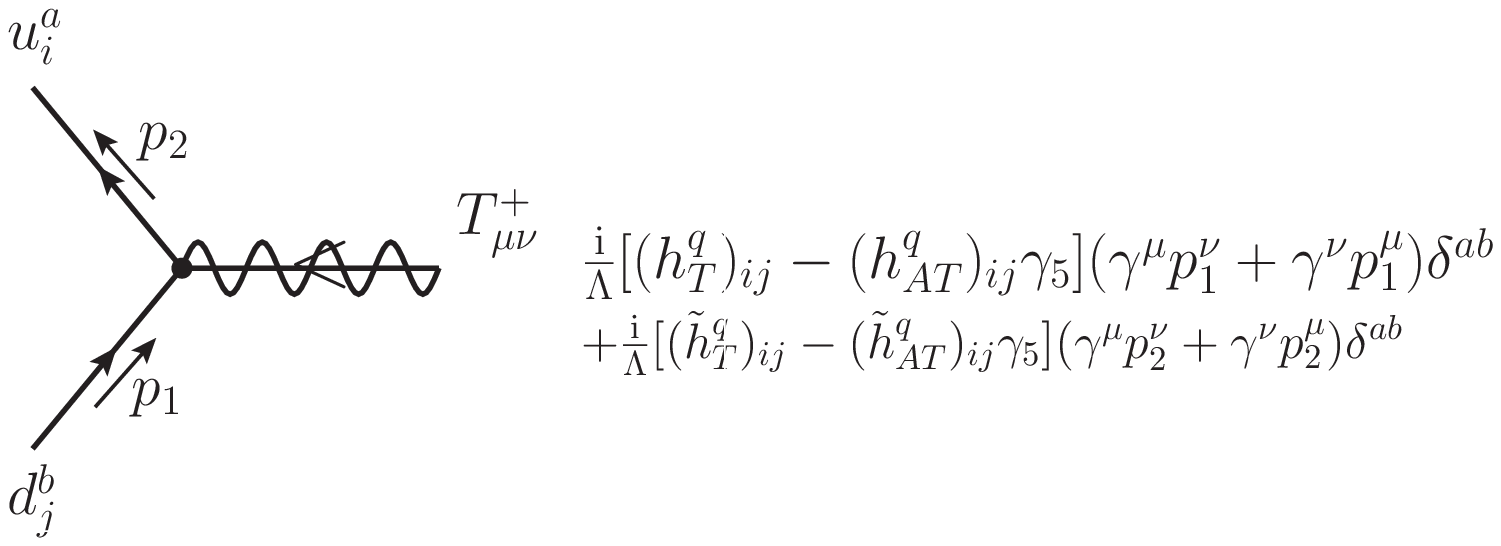}&\hspace{0.5cm}
\includegraphics[scale=.5]{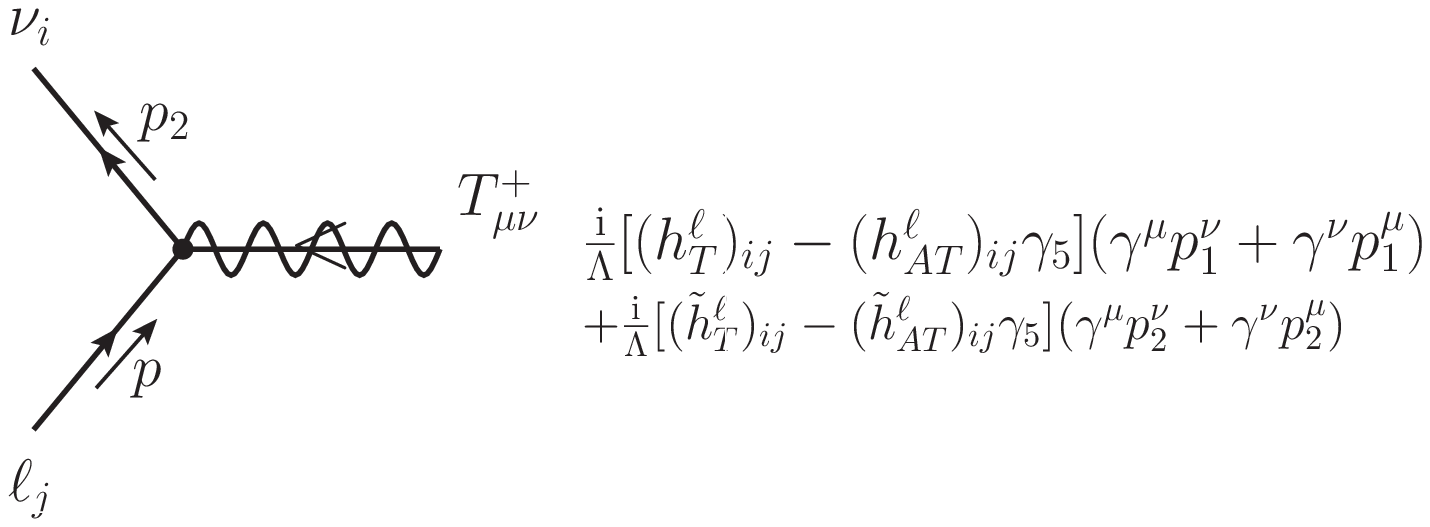}
\end{tabular}
\caption{\label{fig:V'} Interaction vertices between the new charged bosons and SM fermions.}
\end{center}
\end{figure*}

\begin{figure*}[hptb]
\begin{center}
\begin{tabular}{ll}
\includegraphics[scale=.5]{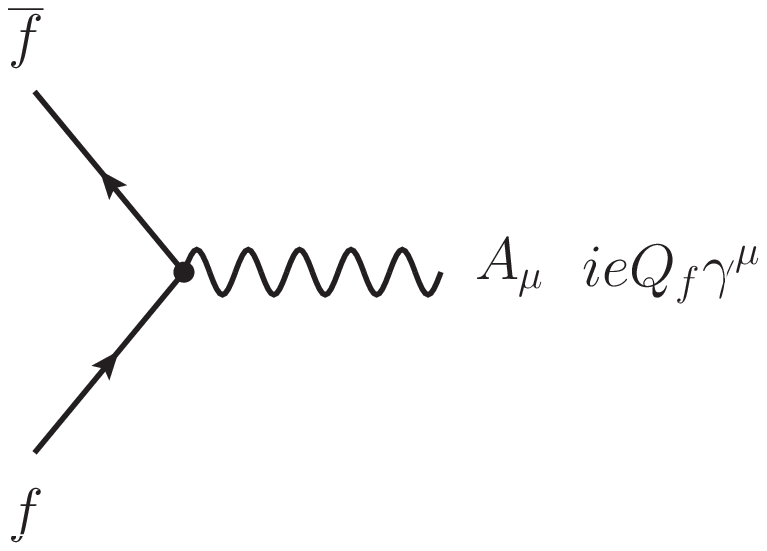}&\hspace{1.0cm}
\includegraphics[scale=.5]{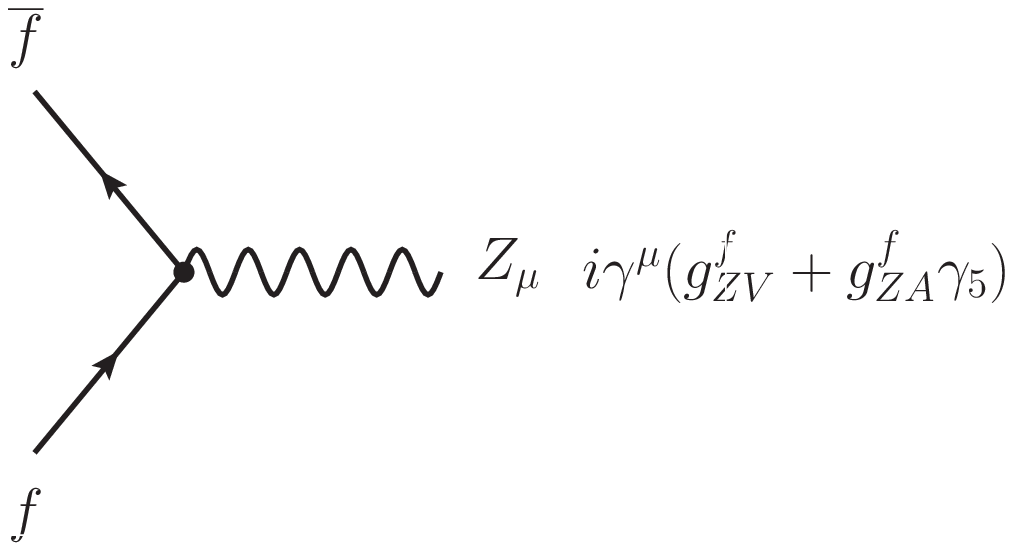}\\
\includegraphics[scale=.5]{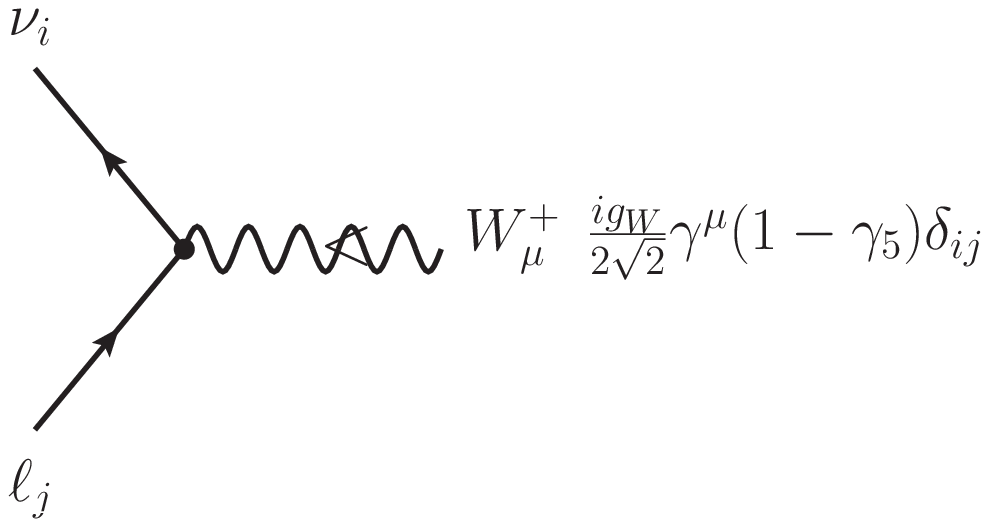}&\hspace{0.5cm}
\includegraphics[scale=.5]{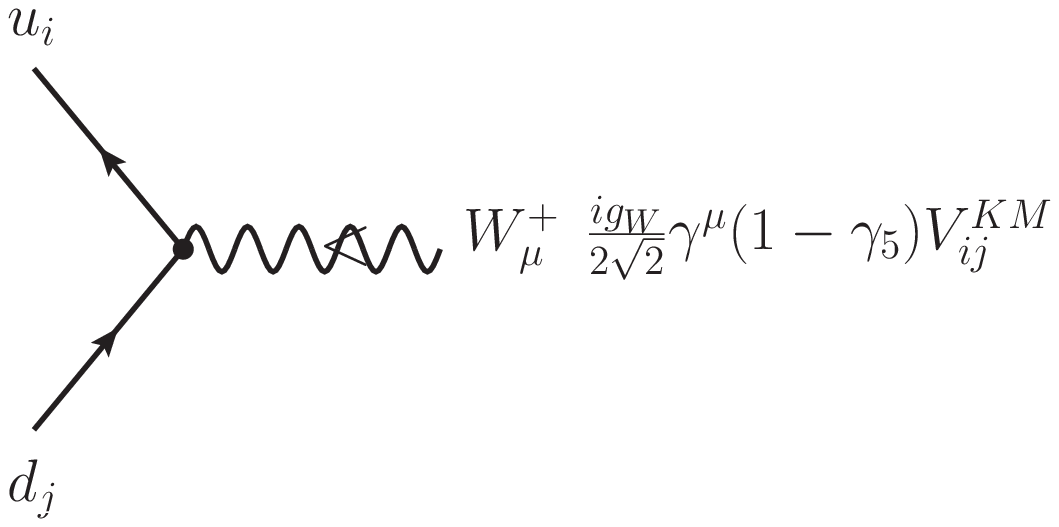}\\
\end{tabular}
\caption{\label{SM_vertex} Interaction vertices between the SM gauge bosons and fermions, where the weak coupling $g_W$ is related to the Fermi constant $G_F$ by $\frac{G_F}{\sqrt{2}}=\frac{g^2_W}{8M^2_W}$.}
\end{center}
\end{figure*}

In this appendix, we provide the Feynman rules used in the analyses.
The interaction vertices between the neutral resonances and the SM
fields are listed in Fig.~\ref{fig:V}.
The interaction vertices between the charged resonances and the SM
fermions are listed in Fig.~\ref{fig:V'}.
The relevant SM vertices are reproduced in Fig.~\ref{SM_vertex} to fix our convention.

%
%
\section{\label{app:FeynRules}FeynRules Implementation}
We implemented our formulation using the FeynRules Mathematica package \cite{Christensen:2008py}.  We will make the FeynRules model files available through the FeynRules model database where those who are interested can download and use it.  We briefly describe this implementation in this section.

This implementation uses two FeynRules model files.  The first is the SM that comes with FeynRules (``SM.fr"), although we will include a copy of this file in case changes are made to the SM model files later.  We added all the new particles, parameters and Lagrangian terms to a new file we call ``plus.fr".

The names of the particles along with their charge, mass and width are listed in Table \ref{tab:FRparticles}.
\begin{table}[!tbh]
\begin{center}
\begin{tabular}{|l|lll|}
\hline
& Name & Mass & Width\\
\hline\hline
$S$ & SV & MSV & WSV\\
$V_\mu$ & VV & MVV & WVV\\
$T_{\mu\nu}$ & TV & MTV & WTV\\
$S^\pm$ & SVP$\pm$ & MSVP & WSVP\\
$V^\pm_\mu$ & VVP$\pm$ & MVVP & WVVP\\
$T^\pm_{\mu\nu}$ & TVP$\pm$ & MTVP & WTVP\\
\hline
\end{tabular}
\end{center}
\caption{\label{tab:FRparticles}Particles implemented in FeynRules.  In the first column is the symbol we use for the particle in this article.  In the last 3 columns are the ASCII names we use for these particles, their masses and widths in FeynRules.  These are the names that would be used in a simulation.  The masses are set to 1~TeV by default while the widths are set to be 20~GeV by default, but both of these parameters are free and can be set by the user.}
\end{table}
These are the names that are used when running a Monte Carlo program.

The couplings are implemented with names that are similar to the ones we use in this article.  However, there is a six-character limit in CalcHEP for the names of parameters, so some names are shortened.  We list the parameter names in Table \ref{tab:FRparameters}.
\begin{table}[!tbh]
\begin{center}
\begin{tabular}{|l|lcc|}
\hline
& Name & Real part & Imaginary part\\
\hline\hline
$g^f_{Sij}$ & gSfij & gSfRij & gSfIij\\
$g^f_{Pij}$ & gPfij & gPfRij & gPfIij\\
$g^g_S$ & gSg &&\\
$g^f_{Vij}$ & gVfij & gVfRij & gVfIij\\
$g^f_{Aij}$ & gAfij & gAfRij & gAfIij\\
$g^g_V$ & gVg &&\\
$g^f_{Tij}/\Lambda$ & gTfij & gTfRij & gTfIij\\
$g^f_{ATij}/\Lambda$ & gUfij & gUfRij & gUfIij\\
$g^g_T/\Lambda$ & gTg &&\\
\hline
$h^f_{Sij}$ & hSfij & hSfRij & hSfIij\\
$h^f_{Pij}$ & hPfij & hPfRij & hPfIij\\
$h^f_{Vij}$ & hVfij & hVfRij & hVfIij\\
$h^f_{Aij}$ & hAfij & hAfRij & hAfIij\\
$h^f_{Tij}/\Lambda$ & hTfij & hTfRij & hTfIij\\
$h^f_{ATij}/\Lambda$ & hUfij & hUfRij & hUfIij\\
$\tilde{h}^f_{Tij}/\Lambda$ & hYfij & hYfRij & hYfIij\\
$\tilde{h}^f_{ATij}/\Lambda$ & hZfij & hZfRij & hZfIij\\
\hline
\end{tabular}
\end{center}
\caption{\label{tab:FRparameters}Couplings implemented in FeynRules.  In the first column is the symbol we use for the coupling in this article.  In the second column is the ASCII name we use for this coupling in FeynRules while the third and fourth columns contain the the real and imaginary parts.
The letter ``f" refers to the flavor and runs over ``u", ``d" and ``l" for the neutral current couplings while it runs over ``q" and ``l" for the charged current couplings.  The $i$ and $j$ stand for the generation.
}
\end{table}
The couplings are implemented in a very general way in the FeynRules file and are all set to zero by default so that the user can turn on the vertices that they are interested in.  Some are removed from the Monte Carlo file using ``Definitions" in the FeynRules model file.  These can easily be turned on by removing the appropriate definitions (the full couplings are included in the Lagrangians), but the user should remember the hermiticity requirements on some of the couplings.  Furthermore, if the user desires to use this implementation with a Monte Carlo package that can handle complex parameters, it is possible to turn off the splitting of these couplings into real and imaginary parts by commenting out the ``Definitions" line of the complex couplings.

The full Lagrangians described in this article were included in the FeynRules model files.  FeynRules was then run on this file and all the resulting vertices were checked against our independent calculations of these vertices and agreement was found.

The CalcHEP interface \cite{Christensen:2009jx} was then run on this model, which generated a set of CalcHEP model files.  We used CalcHEP to generate the analytic formulas for the squared matrix element for all the $2\rightarrow2$ Drell-Yan processes where we only included the first generation fermions.  We checked these formulas against our own independent calculations of these squared matrix elements and found agreement.  We then used these model files to do the numerical studies described in this article.

These model files are intended to be used in unitary gauge and below the effective cutoff of these vertices.  Furthermore, these interactions are not ultraviolet complete and should not be used for other processes than the Drell-Yan processes described here.

%
%
\section{$d^j_{m,m'}$ Function Review\label{app:d functions}}
In this appendix, we briefly review the Wigner $d^j_{m,m'}$ functions.  Suppose the incoming state has total angular momentum $j$ and total spin $m$ along the direction of motion of one of the particles.  We will call this the $z$-direction and the total spin is the difference of the particles' helicities\footnote{Although orbital angular momentum plays a role, it's $z$-component is 0.  As a result, the total angular momentum $z$-component is equal to the spin $z$-component ($j_z=m$).}.  The final state has the same total angular momentum and spin $m'$ along the direction of motion of one of the final state particles and at an angle $\theta$ with respect to the $z$-direction.  We will define the $x$-direction such that the whole process occurs in the $x$-$z$ plane.  The spin along this direction is the difference of helicity of the final state particles.  With these definitions, the $d^j_{m,m'}$ function is defined as the overlap between the incoming and outgoing states
\begin{equation}
d^j_{m,m'}(\theta) = \left<j,m',\theta|j,m\right> = \left<j,m'\left|e^{iJ_y\theta}\right|j,m\right>~,
\end{equation}
where we have extracted the angular dependence into the operator $\exp(iJ_y\theta)$ which rotates the final state around the $y$-axis to an angle $\theta$ with respect to the $z$-axis.
Commonly used functions are listed in Table~\ref{table:d-functions}.
 For $j=0$, the generator of rotations around the $y$-axis is $J_y=0$ and so
\begin{equation}
d^0_{0,0}(\theta)=1.
\end{equation}
For $j=1$, the generator $J_y$ is
\begin{equation}
J_y = \frac{-i}{\sqrt{2}}\left(\begin{array}{ccc}
0&1&0\\
-1&0&1\\
0&-1&0
\end{array}\right) .
\end{equation}
We also find that
\begin{equation}
J_y^2 = \frac{-1}{2}\left(\begin{array}{ccc}
-1&0&1\\
0&-2&0\\
1&0&-1
\end{array}\right)
\end{equation}
and
\begin{equation}
J_y^{2n} = J_y^2 \quad, \quad J_y^{2n+1}=J_y ,
\end{equation}
where $n>0$.
With this information, $\exp(iJ_y\theta)$ can be expanded and it can be shown that
\begin{equation}
d^1_{m,m'}=e^{iJ_y\theta}=\left(\begin{array}{ccc}
\frac{1}{2}\left(1+\cos\theta\right)&\frac{1}{\sqrt{2}}\sin\theta&\frac{1}{2}\left(1-\cos\theta\right)\\
-\frac{1}{\sqrt{2}}\sin\theta&\cos\theta&\frac{1}{\sqrt{2}}\sin\theta\\
\frac{1}{2}\left(1-\cos\theta\right)&\frac{1}{\sqrt{2}}\sin\theta&\frac{1}{2}\left(1+\cos\theta\right)
\end{array}\right)~,
\end{equation}
where $m$ and $m'$ refer to the elements of the matrix.


\begin{table}
\begin{center}
\renewcommand{\arraystretch}{1.5}
\begin{tabular}{|ccclc|}
\hline
&$d^0_{0,0}$ &$=$& $1$ & \\
&$d^1_{1,\pm1}$ &$=$& $\frac12 (1 \pm \cos\theta)$ & \\
&$d^2_{1,\pm1}$ &$=$& $\frac12 (1 \pm \cos\theta) (2 \cos\theta \mp 1)$ & \\
&$d^2_{2,\pm1}$ &$=$& $- \frac12 (1 \pm \cos\theta) \sin\theta$ & \\
\hline
\end{tabular}
\end{center}
\caption{
\label{table:d-functions}
Wigner $d^j_{m,m'}$ functions relevant for our processes and used in the helicity amplitudes given in Tables~\ref{tab:helam_neutral} and \ref{tab:helam_charged}.  
}
\end{table}


\end{document}